\documentclass[11pt]{article}

\title{Learning Hybrid Control Barrier Functions from Data}

\usepackage{amsmath}
\usepackage{amssymb}
\usepackage{tikz}
\usepackage{pgfplots}
\usepackage{graphicx}
\usepackage{epstopdf}
\usepackage[labelfont=bf,font=footnotesize]{caption}
\usepackage[labelfont=bf,font=footnotesize]{subcaption}
\usepackage{xcolor}
\allowdisplaybreaks  
\usepackage{cite}
\usepackage{algorithmicx}
\usepackage{algpseudocode}
\usepackage{algorithm}
\usepackage{bbm}
\usepackage{makecell}
\usepackage{authblk}
\usepackage{fullpage,etoolbox}
\usepackage{hyperref}

\allowdisplaybreaks

\usepackage{amsthm}

\newtheorem{definition}{Definition} 
\newtheorem{theorem}{Theorem} 
\newtheorem{proposition}{Proposition} 

\newtheorem{lemma}{Lemma}
\newtheorem{example}{Example}

\usepackage{enumitem}
\setlist{nosep}

\author[1]{Lars Lindemann}
\author[2]{Haimin Hu}
\author[2]{Alexander Robey}
\author[2]{Hanwen Zhang}
\author[1]{Dimos V.\ Dimarogonas}
\author[3]{Stephen Tu}
\author[2]{Nikolai Matni}
\affil[1]{Division of Decision and Control Systems,  KTH Royal Institute of Technology}
\affil[2]{Department of Electrical and Systems Engineering, University of Pennsylvania}
\affil[3]{Google Brain Robotics}

\date{\today}

\begin{document}



\maketitle

\begin{abstract}
Motivated by the lack of systematic tools to obtain safe control laws for hybrid systems, we propose an optimization-based framework for learning certifiably safe control laws from data. In particular, we assume a setting in which the system dynamics are known and in which data exhibiting safe system behavior is available. We propose hybrid control barrier functions for hybrid systems 
as a means to synthesize safe control inputs. Based on this notion, we present an optimization-based framework to learn such hybrid control barrier functions from data. Importantly, we identify sufficient conditions on the data such that feasibility of the optimization problem ensures correctness of the learned hybrid control barrier functions, and hence the safety of the system. We illustrate our findings in two simulations studies, including a compass gait walker. 
\end{abstract}

\section{Introduction}
\label{sec:introduction}

Consider the following \emph{safety-critical} scenarios: autonomous vehicles in urban areas \cite{schwarting2018planning}, 
exoskeletons for improving mobility of lower-body impaired users \cite{tucker2019preference}, and robots navigating a warehouse using semantic logic \cite{kress2009temporal}. 
These systems are all described by \emph{hybrid dynamics}, i.e., states and transitions are both continuous and discrete, due to either their physics, or to higher level logical decision making and importantly, 
share 
that: 1) \emph{data} exhibiting safe behavior is readily available or easily collected, and 2) in most cases, their hybrid system dynamics are well understood and can be identified. Based on these observations, we propose an optimization-based approach to learning \emph{provably safe controllers} for hybrid systems using \emph{hybrid control barrier functions} (HCBF).

\textbf{Related work:} \emph{Barrier functions} (BFs) were introduced in~\cite{prajna2007framework} to certify the safety of continuous-time systems. Nonsmooth~\cite{glotfelter2017nonsmooth} and hybrid~\cite{glotfelter2019hybrid} BFs have also been defined. Different to our work, these focus on BFs with jumps for discontinuous systems.  \emph{Control barrier functions} (CBFs) for continuous-time control systems  appeared in~\cite{wieland2007constructive} for synthesizing feedback control laws that ensure safety by enforcing forward invariance of a desired safe set.  Reciprocal~\cite{ames2014control} and zeroing~\cite{ames2017control} CBFs were proposed as less restrictive alternatives that do not enforce  forward invariance on subsets of the safe set. Such CBFs can be used as a safety guard via convex quadratic programs \cite{ames2014control,ames2017control}, a feature that has been used for safe learning \cite{khojasteh2019probabilistic,taylor2020control}. CBFs for discrete-time control systems can be found in~\cite{agrawal2017discrete,ohnishi2019barrier,cavorsi2020tractable}.
Related to this paper, \emph{hybrid BFs} were proposed in \cite{maghenem2019characterizations,bisoffi2018hybrid} in the hybrid systems framework of~\cite{goebel2012hybrid} as a means of certifying the safety of hybrid systems. Analogous hybrid BFs are introduced in~\cite{prajna2007framework} for the hybrid automata modeling framework of~\cite{lygeros2003dynamical}.

The underlying challenge and bottleneck in ensuring safety using (control) BFs is the construction of these functions.  In~\cite{prajna2007framework}, the authors propose a sum-of-squares programming approach for finding polynomial (hybrid) barrier functions for polynomial (hybrid) systems and semi-algebraic sets. For finding CBFs, bilinear sum-of-squares programming and analytic approaches are proposed in~\cite{ames2019control,xu2017correctness,wang2018permissive}; such methods, however, only apply to a restrictive class of systems and have limited scalability.  Recent approaches attempt to circumvent these limitations by treating the CBF synthesis task as a machine learning problem. In~\cite{yaghoubi2020training}, a deep neural network is trained to imitate the control law obtained from a CBF, whereas in~\cite{srinivasan2020synthesis}, a CBF is synthesized from safe and unsafe data samples using support vector machines. In \cite{saveriano2019learning}, the authors cluster observed data and learn a linear CBF for each cluster.  While the aforementioned works \cite{yaghoubi2020training,srinivasan2020synthesis,saveriano2019learning} present detailed empirical validation of their methods, no formal correctness proofs are provided. In~\cite{jin2020neural}, a Lyapunov, barrier, and
a policy function is learned from data: the validity of the learned certificates are then verified post-hoc using Lipschitz arguments. In~\cite{ferlez2020shieldnn}, a method is proposed that learns a provably correct neural net safety guard for kinematic bicycle models.  
Finally, \cite{robey2020learning} propose a data-driven approach for learning CBFs for smooth nonlinear systems assuming known Lipschitz dynamics, as well as availability of expert demonstrations illustrating safe behavior. Sufficient conditions ensuring correctness of the learned CBF using a Lipschitz argument are also given.

\textbf{Contributions: } For a class of hybrid control systems, we define \emph{hybrid control BFs} as a means to enforce forward invariance of a safe set. We provide sufficient conditions in terms of a HCBF that, if satisfied by a control law, ensure safety. We then show that learning HCBFs from data can be cast as a constrained optimization problem, and provide conditions under which a feasible solution is a valid HCBF.  Finally, we present simulations showcasing the benefits of our framework.

\section{Preliminaries and Problem Formulation}
\label{sec:backgound}


\emph{Notation: }Let $\text{dom}(z):=\{(t,j)\in \mathbb{R}_{\ge 0}\times\mathbb{N}| \exists \zeta\in\mathbb{R}^{n_z} \text{ s.t. } z(t,j)=\zeta\}$ be the domain of a function $z:\mathbb{R}_{\ge 0}\times\mathbb{N}\to  \mathbb{R}^{n_z}$. A function $\alpha:\mathbb{R}\to\mathbb{R}$ is an extended class $\mathcal{K}$ function if $\alpha$ is strictly increasing and $\alpha(0)=0$. For $\epsilon>0$ and $p\ge 1$, let $\mathcal{B}_{\epsilon,p}(z^i):=\{z\in\mathbb{R}^{n_z}\, \big{|}\, \|z-z^i\|_p\le \epsilon\}$ be the closed $p$-norm ball around $z^i\in\mathbb{R}^{n_z}$. Let bd$(\mathcal{C})$ and int$(\mathcal{C})$ be the boundary and interior of a set $\mathcal{C}$. 

\paragraph{Control Barrier Functions}
\label{sec:cbf}
At time $t\in\mathbb{R}_{\ge 0}$, let $x(t)\in\mathbb{R}^n$ be the state  of the dynamical control system  described by the initial value problem
\begin{align}\label{eq:system}
\dot{x}(t)=f(x(t))+g(x(t))u(x(t)), \; x(0)\in\mathbb{R}^n
\end{align}
where $f:\mathbb{R}^n\to\mathbb{R}^n$ and $g:\mathbb{R}^n\to\mathbb{R}^{n\times m}$ are continuous functions. Let the solutions to \eqref{eq:system} under a continuous control law $u:\mathbb{R}^n\to\mathbb{R}^m$ be $x:\mathcal{I}\to \mathbb{R}^n$ where $\mathcal{I}\subseteq\mathbb{R}_{\ge 0}$ is the maximum definition interval of $x$. We do not assume forward completeness of \eqref{eq:system} under $u$ here, i.e., $\mathcal{I}$ may be bounded.

Consider next a continuously differentiable function $h:\mathbb{R}^n\to\mathbb{R}$ and define the set $\mathcal{C}:=\{x\in\mathbb{R}^n \, \big{|} \, h(x)\ge 0\},$ which defines a set that we wish to certify as safe, i.e., that it satisfies prescribed safety specifications and can be made forward invariant through an appropriate choice of control action. Note that $\mathcal{C}$ is closed and further assume that $\mathcal{C}$ is not the empty set. Now, let $\mathcal{D}$ be an open set that is such that $\mathcal{D}\supseteq \mathcal{C}$. The function $h(x)$ is said to be a \emph{valid control barrier function} on $\mathcal{D}$ if there exists a locally Lipschitz continuous extended class $\mathcal{K}$ function $\alpha:\mathbb{R}\to\mathbb{R}$ such that 
\begin{align*}
\sup_{u\in \mathcal{U}} 
\langle \nabla h(x), f(x)+g(x)u\rangle \ge -\alpha(h(x))
\end{align*} 
holds for all $x\in\mathcal{D}$, where $\mathcal{U}\in\mathbb{R}^n$ defines constraints on the control input $u$. Consequently, we define the set of \emph{CBF consistent inputs} induced by a valid CBF $h(x)$ to be 
\begin{align*}
K_{\text{CBF}}(x):=\{u\in\mathbb{R}^m \, \big{|} \, 
\langle \nabla h(x), f(x)+g(x)u\rangle \ge -\alpha(h(x))\}.
\end{align*}
The next result follows mainly from \cite{ames2017control}.\footnote{We provide a slightly modified version to account for  \cite[Remark 5]{ames2019control} and to not require that $\nabla h(x)\neq 0$ when $x\in \text{bd}(\mathcal{C})$ by using the Comparison Lemma \cite[Lemma 3.4]{Kha96}.}
\begin{lemma}\label{lem:1}
Assume that $h(x)$ is a valid control barrier function on $\mathcal{D}$ and that $u:\mathcal{D}\to \mathcal{U}$ is a continuous function with $u(x)\in K_{\text{CBF}}(x)$. Then $x(0)\in\mathcal{C}$ implies $x(t)\in\mathcal{C}$ for all $t\in \mathcal{I}$. If $\mathcal{C}$ is compact, it follows that $\mathcal{C}$ is forward invariant under $u(x)$, i.e., $\mathcal{I}=[0,\infty)$.

\begin{proof}
First note that $\dot{v}(t)=-\alpha(v(t))$ with $v(0)\ge 0$ admits a unique solution $v(t)$ that is such that $v(t)\ge 0$ for all $t\ge 0$ \cite[Lemma 4.4]{Kha96}. Each solution $x(t)$ to \eqref{eq:system} under $u(x)$  is now, due to the chain rule and since $u(x)\in K_{\text{CBF}}(x)$, such that $\dot{h}(x(t))\ge -\alpha(h(x(t)))$ for all $t\in\mathcal{I}$. Using the Comparison Lemma \cite[Lemma 3.4]{Kha96} and assuming that $h(x(0))\ge 0$, it follows that $h(x(t))\ge v(t)\ge 0$ for all $t\in\mathcal{I}$, i.e., $x(0)\in\mathcal{C}$ implies $x(t)\in\mathcal{C}$ for all $t\in \mathcal{I}$. Note next that \eqref{eq:system} is defined on $\mathcal{D}$ since $u(x)$ is only defined for $x\in \mathcal{D}$. Since $x\in \mathcal{C}$ for all $t\in\mathcal{I}$ and when $\mathcal{C}$ is compact, it follows by \cite[Theorem 3.3]{Kha96} that $\mathcal{I}=[0,\infty)$, i.e., $\mathcal{C}$ is forward invariant.
\end{proof}
\end{lemma}

\paragraph{Hybrid Systems }
\label{sec:RRSTL}

We model and analyze hybrid systems using the formalism of~\cite{goebel2012hybrid,goebel2009hybrid}.

\begin{definition}\label{def:hybrid_system}
	A hybrid system \cite{goebel2012hybrid} is a tuple $\mathcal{H}:=(C,F,D,G)$ where $C\subseteq\mathbb{R}^{n_z}$, $D\subseteq\mathbb{R}^{n_z}$, $F:\mathbb{R}^{n_z}\to \mathbb{R}^{n_z}$, and $G:\mathbb{R}^{n_z}\to\mathbb{R}^{n_z}$ are the flow and jump sets and the continuous flow and jump maps, respectively. At the hybrid time $(t,j)\in \mathbb{R}_{\ge 0}\times\mathbb{N}$, let $z(t,j)\in \mathbb{R}^{n_z}$ be the hybrid state with initial condition $z(0,0)\in C\cup D$ and the hybrid system dynamics
	\begin{align}\label{eq:hybrid_systems}
	\begin{cases}
	\dot{z}(t,j)= F(z(t,j)) \ &\text{  for  } \ z(t,j) \in C,\\ 
	 z(t,j+1)= G(z(t,j)) \ &\text{  for  } \ z(t,j)  \in D.
	\end{cases}
	\end{align}
\end{definition}

 Solutions to \eqref{eq:hybrid_systems} are parameterized by $(t,j)$, where $t$ indicates continuous \emph{flow} according to $F(z)$ and $j$ indicates discontinuous \emph{jumps} according to $G(z)$. Now let $\mathcal{E} \subseteq \mathbb{R}_{\ge 0}\times \mathbb{N}$ be a \emph{hybrid time domain} \cite[Ch. 2.2]{goebel2012hybrid}, i.e., $\mathcal{E}$ is an infinite union of intervals of the form $[t_j,t_{j+1}]\times \{j\}$ or a finite union of intervals of the form $[t_j,t_{j+1}]\times \{j\}$ where the last interval, if it exists, has the form $[t_j,t_{j+1}]\times \{j\}$, $[t_j,t_{j+1})\times \{j\}$, or $[t_j,\infty)\times \{j\}$. We now formally define a \emph{hybrid solution} $z:\mathcal{E}\to C\cup D$ to $\mathcal{H}$.

\begin{definition}\label{def:hybrid_solutions}
	A function $z:\mathcal{E}\to  C \cup D$ is a hybrid solution to $\mathcal{H}$ if $z(0,0)\in C \cup D$ and 
	\begin{itemize}[leftmargin=*]
	    \item for each $j\in\mathbb{N}$ such that $I_j:= \{t\in\mathbb{R}_{\ge 0}|(t,j)\in \text{dom}(z) \}$ is not a singleton, $z(t,j)\in C$ and  $\dot{z}(t,j)=F(z(t,j))$ for all $t\in[\min I_j, \sup I_j)$
	    \item for each $(t,j)\in\text{dom}(z)$ s.t. $(t,j+1)\in\text{dom}(z)$, $z(t,j)\in D$ and $z(t,j+1)=G(z(t,j))$.
	\end{itemize}
\end{definition}

\begin{example}\label{ex:1}
Consider the bouncing ball example from \cite[Example 1.1]{goebel2012hybrid}. Let $x\in\mathbb{R}$ be the (vertical) position and $v\in\mathbb{R}$ be the (vertical) velocity of a point mass. Let $z:=\begin{bmatrix}x & v \end{bmatrix}^T$ and 
\begin{align*}
\dot{z}(t,j)= \begin{bmatrix} v(t,j) & - \rho \end{bmatrix}^T \ &\text{ for } \ z(t,j) \in C,\\ 
z(t,j+1)= \begin{bmatrix} x(t,j) & - \kappa v(t,j) \end{bmatrix}^T \ &\text{ for  } \ z(t,j) \in D,
\end{align*}
where $\rho>0$ (gravity), $\kappa\in(0,1)$ (damping), $C:=\{z\in\mathbb{R}^2|x>0 \text{ or } x=0,v\ge 0\}$, and $D:=\{z\in\mathbb{R}^2|x=0,v<0\}$. For an initial condition $z(0,0):=\begin{bmatrix}x(0,0) & v(0,0) \end{bmatrix}^T$ with $x(0,0)> 0$, the hybrid time domain $\mathcal{E}$ is an infinite union of intervals of the form $[t_j,t_{j+1}]\times \{j\}$ where, as $j\to \infty$, $t_j$ converges to some constant value and where $\|z(t,j)\|$ converges to zero. 
\end{example}

\paragraph{Problem Formulation}
The class of \emph{hybrid control systems} that we consider is
\begin{align}\label{eq:hybrid_control_systems}
\begin{cases}
\dot{z}(t,j)= f_c(z(t,j))+g_c(z(t,j))u_c(z(t,j)) & \ \text{ for  }\ z(t,j) \in C  \\
z(t,j+1)= f_d(z(t,j))+g_d(z(t,j))u_d(z(t,j)) & \ \text{ for  }\ z(t,j) \in D
\end{cases}
\end{align}
where $u_c:C \to \mathcal U_c \subseteq \mathbb{R}^{m_c}$ and $u_d:D \to \mathcal U_d \subseteq\mathbb{R}^{m_d}$ are continuous control laws  during flows and jumps, respectively. Let $f_c:\mathbb{R}^{n_z}\to\mathbb{R}^{n_z}$, $f_d:\mathbb{R}^{n_z}\to\mathbb{R}^{n_z}$, $g_c:\mathbb{R}^{n_z}\to\mathbb{R}^{n_z\times n_c}$, and $g_d:\mathbb{R}^{n_z}\to\mathbb{R}^{n_z\times m_d}$ be locally Lipschitz continuous functions.\footnote{We again do not assume completeness of the system \eqref{eq:hybrid_control_systems} under $u_c$ and $u_d$. Completeness here means that the hybrid time domain $\text{dom}(z)$ is unbounded (see \cite[Ch. 2.2]{goebel2012hybrid} for a formal definition).}

We are additionally given a set of \emph{expert trajectories} consisting of $N_c$ and $N_d$ discretely sampled data-points along flows and jumps as 
\begin{align*}
    Z_\text{dyn}^c:=\{(z^i,u^i_c)\}_{i=1}^{N_c} & \ \text{ for } \ z^i\in C,\\
    Z_\text{dyn}^d:=\{(z^i,u^i_d)\}_{i=1}^{N_d} & \ \text{ for } \ z^i\in D,
\end{align*} 
as illustrated in Figure \ref{fig:1} (left).\footnote{We refer to the collection of data points $Z_\text{dyn}^c$ and $Z_\text{dyn}^d$ as expert trajectories  to emphasize that this is a natural way of collecting the $\{(z^i,u_c^i)\}$ and $\{(z^i,u_d^i)\}$ pairs from the system \eqref{eq:hybrid_control_systems}.  We note, however, that our method simply requires a collection of state-action pairs $\{(z^i,u_c^i)\}$ and $\{(z^i,u_d^i)\}$ demonstrating safe behavior.} It is assumed that each $z^i\in\text{int}(\mathcal{S})$ where $\mathcal{S}\subseteq \mathbb{R}^{n_z}$ is the \emph{geometric safe set}, i.e., the set of safe states as naturally specified on a subset of
the system configuration space (e.g., to avoid collision, vehicles must maintain a minimum
separating distance).

Our goal is now to learn, from $Z_\text{dyn}^c$ and $Z_\text{dyn}^d$,  a twice continuously differentiable function $h:\mathbb{R}^{n_z}\to\mathbb{R}$ such that
\begin{align}\label{eq:set_C}
\mathcal{C}:=\{z\in\mathbb{R}^{n_z} \, \big{|} \, h(z)\ge 0\}
\end{align}
is a subset of the geometric safe set $\mathcal{S}$, 
and that $\mathcal{C}$ can be made forward invariant by appropriate control actions $u_c(z)\in \mathcal U_c$ and $u_d(z)\in \mathcal U_d$.


\section{Learning Hybrid Control Barrier Functions from Data} \label{sect:learning-hcbfs}

We begin by defining a suitable notion of \emph{hybrid control barrier functions}, and show that they provide a mechanism for safe control of the hybrid control system \eqref{eq:hybrid_control_systems}.  We then show how such HCBFs can be learned from data via a constrained optimization problem, and provide sufficient conditions under which a feasible solution is a valid HCBF.

\begin{figure*}
\centering
\includegraphics[scale=0.25]{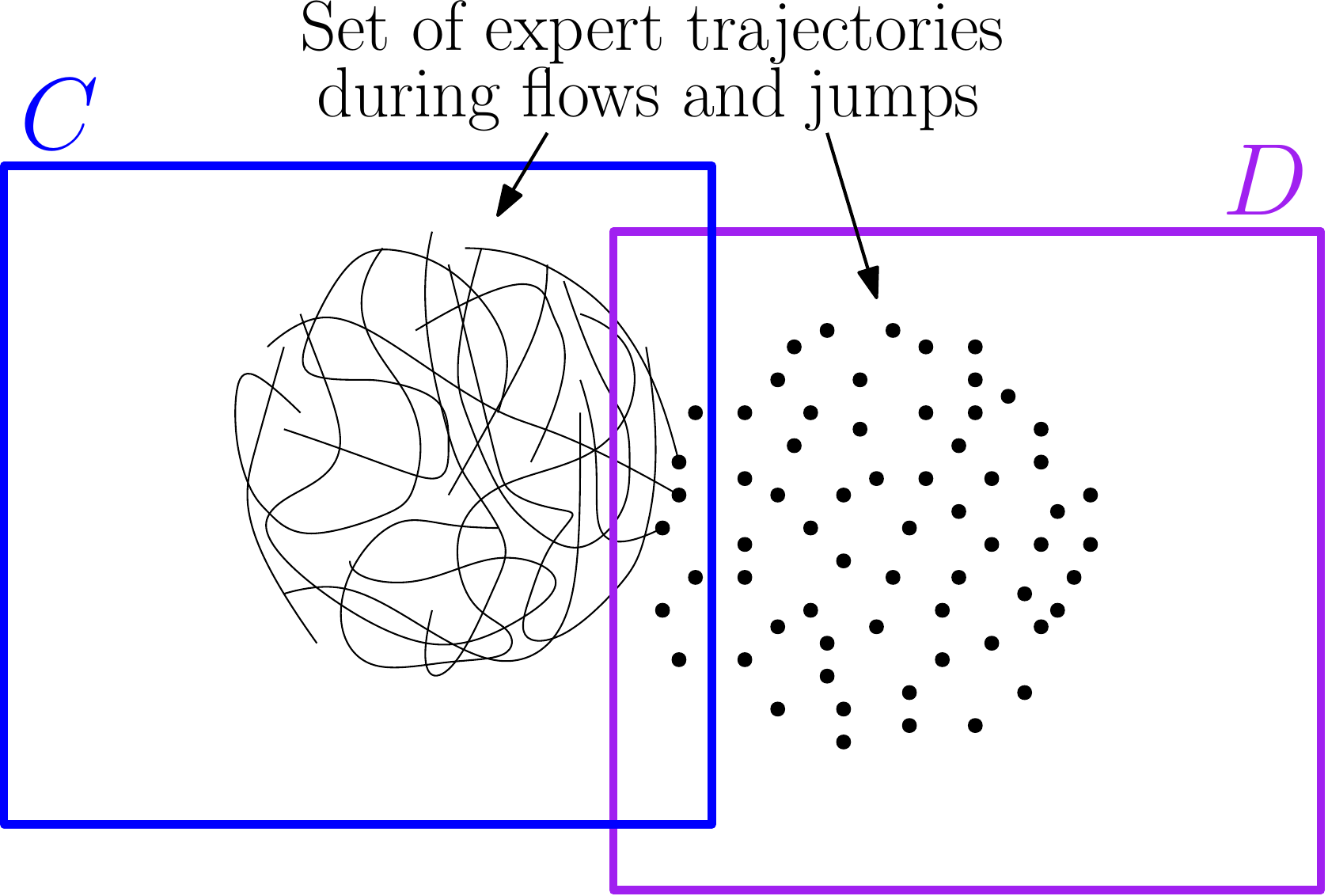}
\hspace{0.5cm}\includegraphics[scale=0.25]{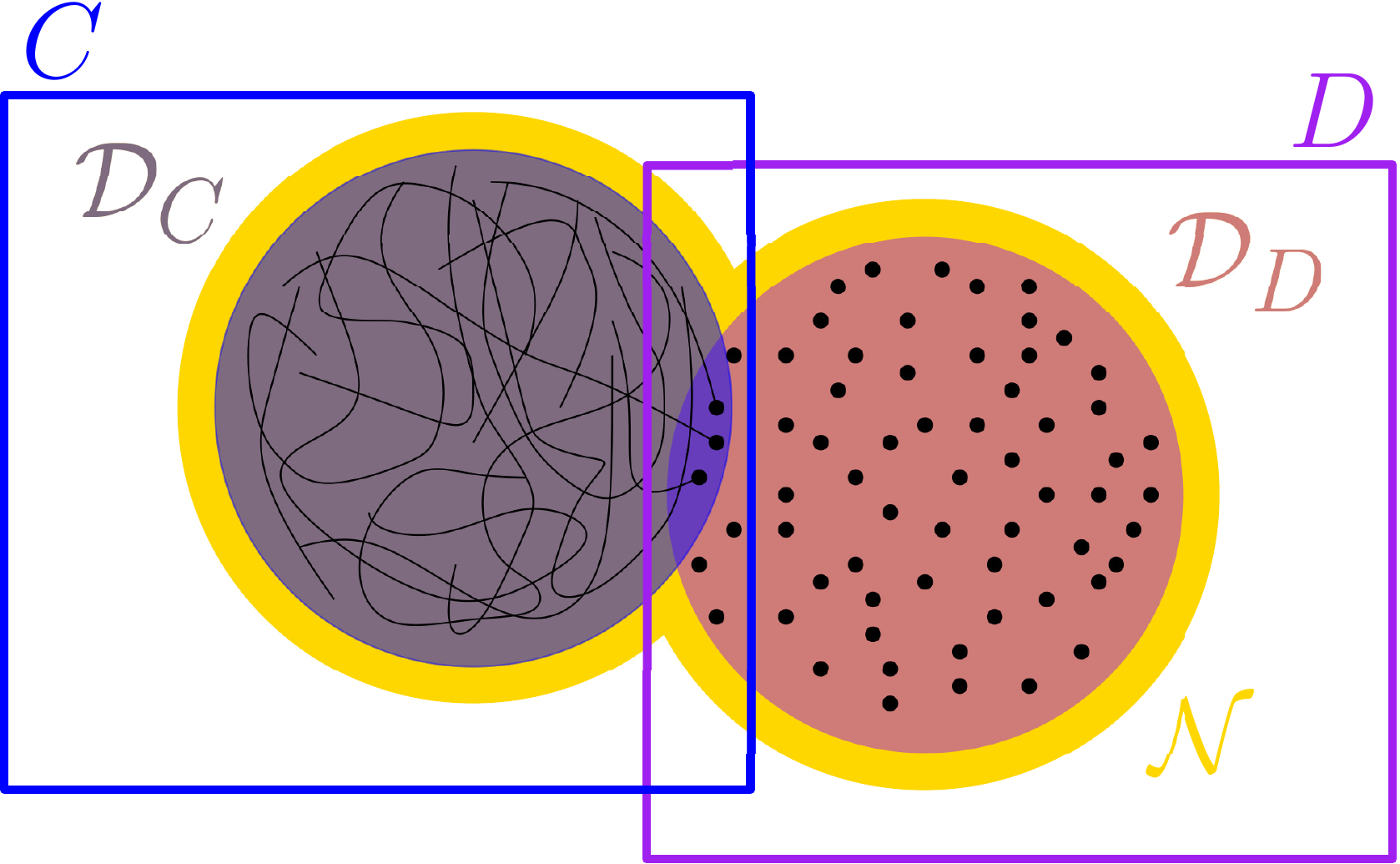}\hspace{0.5cm}\includegraphics[scale=0.25]{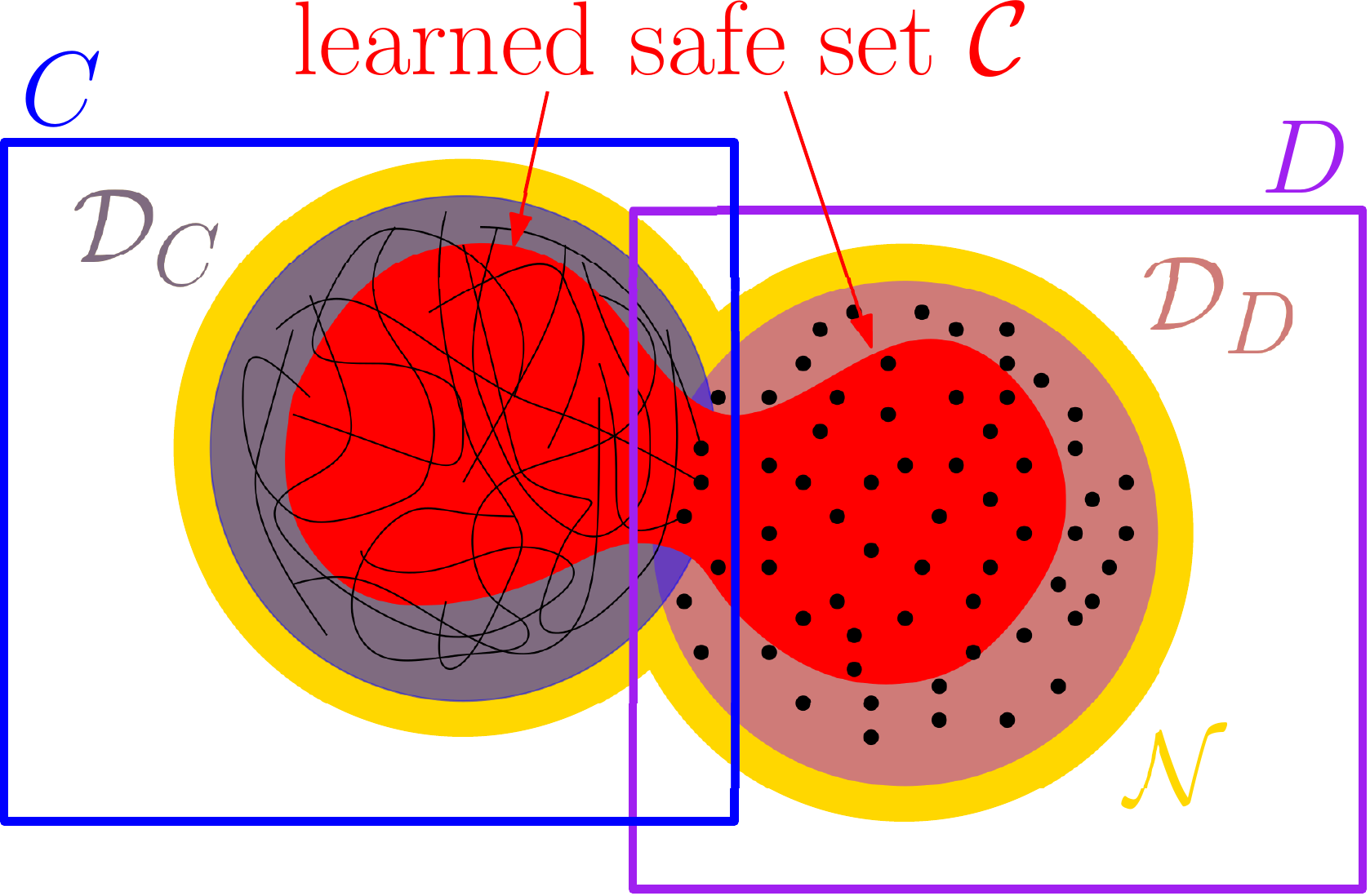}
\caption{Problem setup (left): The flow and jump sets $C$ and $D$ (blue and purple boxes) and the set of safe expert trajectories during flows and jumps (black lines and dots). Set definitions (middle): The sets $\mathcal{D}_C$ and $\mathcal{D}_D$ (black and light red balls) are the union of $\epsilon$ balls around the expert trajectories during flows and jumps. The set $\mathcal{N}$ (golden rings), defined around $\mathcal{D}_C$ and $\mathcal{D}_D$, ensures that the learned safe set $\mathcal{C}$ is such that $\mathcal{C}\subset\mathcal{D}\subseteq\mathcal{S}$. Note that the geometrical safe set $\mathcal{S}$ is not depicted here. Desired result (right): The learned safe set $\mathcal{C}$ (red region) is defined via the learned valid HCBF $h(z)$.}
\label{fig:1}
\end{figure*}

\paragraph{Hybrid Control Barrier Functions}
Let $h:\mathbb{R}^{n_z}\to\mathbb{R}$ be a twice continuously differentiable function for which the set $\mathcal{C}$ in \eqref{eq:set_C} is not empty, and assume that $\mathcal{C}\subseteq C\cup D$. 
Such an assumption is natural as we are only interested in regions where the system \eqref{eq:hybrid_control_systems} is defined. Consider now the sets $\mathcal{D}_C\subseteq C$ and $\mathcal{D}_D\subseteq D$ that are such that 
\begin{align*}
    \mathcal{C}&\cap C \subseteq \mathcal{D}_C,\\
    \mathcal{C}&\cap D \subseteq\mathcal{D}_D
\end{align*}
so that $\mathcal{C}\subseteq \mathcal{D}_C\cup \mathcal{D}_D$\footnote{This follows as we assume that $\mathcal{C}\subseteq C\cup D$, which results in $(\mathcal{C}\cap C)\cup(\mathcal{C}\cap D)=\mathcal{C} \cap (C\cup D)=\mathcal{C}$, and since $(\mathcal{C}\cap C)\cup(\mathcal{C}\cap D)\subseteq \mathcal{D}_C\cup \mathcal{D}_D$ by the choices of the sets $\mathcal{D}_C$ and $\mathcal{D}_D$. }, which ensures that the set $\mathcal{D}:=\mathcal{D}_C\cup\mathcal{D}_D$ fully covers $\mathcal{C}$ -- see Fig.~\ref{fig:1}(middle) and (right). The sets $\mathcal{D}_C$ and $\mathcal{D}_D$ are the equivalent to the set $\mathcal{D}$ in Section \ref{sec:cbf}, but now considered separately for flows and jumps.

\begin{definition}
The function $h(z)$ is said to be a \emph{valid hybrid control barrier function}  on $\mathcal{D}$ if there exists a locally Lipschitz continuous extended class $\mathcal{K}$ function $\alpha:\mathbb{R}\to \mathbb{R}$ such that 
    \begin{align}\label{eq:cbf_const_hybrid}
        \sup_{u_c\in \mathcal{U}_c} 
        \langle \nabla h(z), f_c(z)+g_c(z)u_c\rangle \ge -\alpha(h(z)) \ \text{ for all $z\in \mathcal{D}_C$,}
    \end{align} 
and it holds that 
    \begin{align}\label{eq:cbf_disc_hybrid}
        \sup_{u_d\in \mathcal{U}_d} h(f_d(z)+g_d(z)u_d) \ge 0 \ \text{ for all $z\in \mathcal{D}_D$.}
    \end{align}
\end{definition}
Based on the above definition, we define the sets of HCBF consistent inputs to be
\begin{align*}
K_{\text{HCBF},c}(z)&:=\{u_c\in\mathcal{U}_c\, \big{|} \, 
\langle \nabla h(z), f_c(z)+g_c(z)u_c\rangle \ge -\alpha(h(z))\},\\
K_{\text{HCBF},d}(z)&:=\{u_d\in\mathcal{U}_d\, \big{|} \, 
 h(f_d(z)+g_d(z)u_d) \ge 0\}.
 \end{align*}
  We further assume for the remainder of the paper that the set $\mathcal{D}_C$ is open.\footnote{If $\mathcal{D}_C$ is not open, one can instead assume that $\mathcal{C}\setminus \mathcal{D}_D$ is strictly contained within $\mathcal{D}_C\cup \mathcal{D}_D$.} This will allow us, in the next result, to establish forward invariance of the set $\mathcal{C}$ under control laws $u_c(z)$ and $u_d(z)$  when the set $\mathcal{C}$ is compact. 
\begin{theorem}\label{thm:1}
Assume that $h(z)$ is a valid hybrid control barrier function on $\mathcal{D}$ and that $u_c:\mathcal{D}_C\to \mathcal{U}_c$ and $u_d:\mathcal{D}_D\to \mathcal{U}_d$ are continuous functions with $u_c(z)\in K_{\text{HCBF},c}(z)$ and $u_d(z)\in K_{\text{HCBF},d}(z)$. Then $z(0,0)\in\mathcal{C}$ implies $z(t,j)\in\mathcal{C}$ for all $(t,j)\in \text{dom}(z)$. If $\mathcal{C}$ is compact and satisfies $\mathcal{C}\subseteq C\cup D$, then the set $\mathcal{C}$ is forward invariant under $u_c(z)$ and $u_d(z)$, i.e., $\text{dom}(z)$ is unbounded.

\begin{proof}
During flows with $I_j$ not being a singleton, and if $h(z(\min(I_j),j))\ge 0$, we infer that $h(z(t,j))\ge 0$ for all $t\in[\min I_j, \sup I_j)$  due to \eqref{eq:cbf_const_hybrid} and as in the proof of Lemma \ref{lem:1}. By continuity of $h(z)$ and $z(t,j)$ and since $\mathcal{C}$ is closed, it also holds that $h(z(\sup I_j,j))\ge 0$ if $I_j=[t_j,t_{j+1}]\times \{j\}$, i.e., the right end point is included in $I_j$.  After each jump, it holds that $h(z(t_{j+1},j+1))\ge 0$ as a consequence of \eqref{eq:cbf_disc_hybrid}. Consequently, $z(0,0)\in\mathcal{C}$ implies $z(t,j)\in\mathcal{C}$ for all $(t,j)\in \text{dom}(z)$. 
 
 We next show that  $\mathcal{C}$ is forward invariant under $u_c(z)$ and $u_d(z)$, i.e., $\text{dom}(z)$ is unbounded, if $\mathcal{C}$ is compact and if $\mathcal{C}\subseteq C\cup D$. First recall that $\mathcal{C}\subseteq \mathcal{D}_C\cup \mathcal{D}_D$ due to $\mathcal{C}\subseteq C\cup D$  and since $\mathcal{C}\cap C \subseteq \mathcal{D}_C$ and $\mathcal{C}\cap D \subseteq\mathcal{D}_D$. Assume now that the hybrid solution $z(t,j)$ to the system \eqref{eq:hybrid_control_systems} under control laws $u_c(z)$ and $u_d(z)$ is maximal\footnote{A hybrid solution $z(t,j)$ is maximal if there exists no other hybrid solution $z'(t,j)$ with $\text{dom}(z)\subset \text{dom}(z')$ and with $z(t,j)=z'(t,j)$ for all $(t,j)\in\text{dom}(z)$. } and that the hybrid time domain $\text{dom}(z)$ is bounded. By defining $(T,J):=\sup_{(t,j)} \text{dom}(z)$, we distinguish between the two cases: 1) $z(T,J)\not\in \text{dom}(z)$ and 2) $z(T,J)\in \text{dom}(z)$. 1)  Note that $z(T,J)\not\in \text{dom}(z)$ can only happen when $I_J$ is not a singleton and when $z(t,j)$ has left the set $\mathcal{D}_C$ by flowing without entering $\mathcal{D}_D$, which would enable a successive jump that is not possible since it is assumed that the solution $z(t,j)$ is maximal. Since the set $\mathcal{C}$ is compact and since $\mathcal{D}_C$ is open, there has to exist a time $t<T$ such that $z(t,J)\in \mathcal{D}_C\setminus \mathcal{C} $ according to \cite[Theorem 3.3]{Kha96}, which does not hold since $(t,J)\in \text{dom}(z)$ and since $z(t,J)\in\mathcal{C}$ as shown previously. Since $z(t,j)$ is maximal, it follows by contradiction that $\text{dom}(z)$ is unbounded. 2) If $z(T,J)\in \text{dom}(z)$, it holds that $z(T,J)\in\mathcal{C}$  so that either $z(T,J)\in \mathcal{D}_D\cap \mathcal{C}$ or $z(T,J)\in (\mathcal{D}_C\cap \mathcal{C}) \setminus (\mathcal{D}_D\cap \mathcal{C})$. In the former case, the solution $z(t,j)$ can be extended by a jump. In the latter case, note that $z(T,J)$ is strictly contained within the set $\mathcal{D}_C \cup \mathcal{D}_D$ so that the solution $z(t,j)$ can be extended into $\mathcal{D}_C$ or into $\mathcal{D}_D$ by flowing. By contradiction, it again follows that $\text{dom}(z)$ is unbounded.
\end{proof}
\end{theorem}

\begin{example}\label{ex:2}
Consider again the bouncing ball example from Example \ref{ex:1}, here artificially equipped with continuous and discrete control inputs for illustrative purposes. The dynamics are
\begin{align*}
\dot{z}(t,j)= \begin{bmatrix} v(t,j) & u_c- \rho \end{bmatrix}^T \, &\text{ for } \, z(t,j) \in C, \\ 
z(t,j+1)= \begin{bmatrix} x(t,j) & - u_dv(t,j) \end{bmatrix}^T  \, &\text{ for } \, z(t,j) \in D,
\end{align*}
and the geometric safe set is $\mathcal{S}:=\{z\in\mathbb{R}^2| v^2\le \zeta\}$. The artificial control inputs $u_c$ and $u_d$ can be thought of as controllable wind resistance and floor damping, respectively. Define the candidate HCBF $h(z):=\zeta-v^2$, such that $\mathcal{C}=\mathcal{S}$. Evaluating constraint \eqref{eq:cbf_const_hybrid} with $\alpha(r):=r$ 
yields
$ -2 v u_c+2v\rho \ge  -\zeta+v^2$ and for which we can always find a suitable $u_c\in\mathbb{R}$. Similarly, evaluating constraint \eqref{eq:cbf_disc_hybrid} yields 
$\zeta-(u_dv)^2 \ge 0$ and for which we can, again, always find a suitable $u_d\in\mathbb{R}$.
\end{example}


\paragraph{An Optimization Based Approach}
\label{sec:compute}

First, define the sets 
\begin{align*}
    Z_\text{safe}^c &:= \{ z^i : (z^i, u^i_c) \in Z_\text{dyn}^c \},\\
    Z_\text{safe}^d &:= \{ z^i : (z^i, u^i_d) \in Z_\text{dyn}^d \}.
\end{align*} 
Towards the goal of learning a valid HCBF, we now define, for $\epsilon_c,\epsilon_d>0$ and $p\ge 1$, the data sets
\begin{subequations}\label{eq:set_D}
\begin{align}
    \mathcal{D}_D&:=D\cap\bigcup_{z^i\in Z_\text{safe}^d}\mathcal{B}_{\epsilon_d,p}(z^i),\\ 
    \mathcal{D}_C &:= \mathcal{D}_C' \backslash \mathrm{bd}(\mathcal{D}_C') \,\text{ where }\, \mathcal{D}_C':=C\cap \bigcup_{z^i\in Z_\text{safe}^c}\mathcal{B}_{\epsilon_c,p}(z^i)
\end{align} 
\end{subequations}
that need to be such that  $\mathcal{D}=\mathcal{D}_C\cup \mathcal{D}_D \subseteq \mathcal{S}$, which can be easily achieved even when data-points $z^i$ are close to $\text{bd}(\mathcal{S})$  by adjusting $\epsilon_c$ and $\epsilon_d$ or by omitting $z^i$. Note that the set $\mathcal{D}_C $ is open by definition. For $\sigma>0$, define 
\begin{align*}
    \mathcal{N}:= \{\text{bd}(\mathcal{D})\oplus\mathcal{B}_{\sigma,p}(0)\} \setminus \mathcal{D},
\end{align*}
where $\mathcal{N}$ is a ring of diameter $\sigma$ that surrounds the set $\mathcal{D}$ (see the golden ring in Figure \ref{fig:1}). We will use the set $\mathcal{N}$ to enforce that the value of the learned HCBF $h(z)$ is negative on $\mathcal{N}$ to ensure that the set $\mathcal{C}$ is contained within the set $\mathcal{D}$, which is a necessary condition for $h(z)$ to be valid. Hence, also assume that points 
\begin{align*}
    Z_N = \{z^i\}_{i=1}^{N_u}
\end{align*} 
are sampled from $\mathcal{N}$, i.e., $z^i\in \mathcal{N}$.  Note that only state data $z^i\in Z_N$ are sampled from $\mathcal{N}$, while no inputs $u_c^i$ or $u_d^i$ are needed, i.e., unsafe data can be obtained by gridding or sampling, and does not require visiting unsafe states. While the set $\mathcal{C}$ defined in \eqref{eq:set_C} considers all $z\in \mathbb{R}^{n_z}$ such that $h(z) \geq 0$, we modify this definition slightly by restricting the domain to the set $\mathcal{N} \cup \mathcal{D}$. This is a natural restriction as we are learning a HCBF from data sampled over the domain $\mathcal{N} \cup \mathcal{D}$, and we therefore instead consider learning a valid \emph{local} HCBF $h(z)$  over $\mathcal{D}$ with respect to the set
\begin{align} \label{eq:local_C}
\mathcal{C}:=\{z\in\mathcal{N} \cup \mathcal{D}\, \big{|} \, h(z)\ge 0\}.
\end{align}

\textbf{The optimization problem: }We now propose an optimization problem for learning a valid local HCBF, and then prove its correctness.  We solve
\begin{subequations}\label{eq:opt}
\begin{flalign}
    &\min_{h \in \mathcal{H}} \:\: \|h\|   \: \nonumber \\
    &\mathrm{s.t.}~~ h(z^i) \geq \gamma_\text{safe}, \: \forall z^i \in Z_\text{safe}^c\cup Z_\text{safe}^d \;\;\;\;\;\;\;\;\;\;\;\;\;\;\;\;\;\;\;\;\;\;\;\;\;\;\;\;\;\;\;\; \text{: safe set with margin $\gamma_\text{safe}>0$}\label{eq:cons_1}\\
    &~~~~~~ h(z^i) \leq -\gamma_\text{unsafe}, \: \forall z^i \in Z_N \;\;\;\;\;\;\;\;\;\;\;\;\;\;\;\;\;\;\;\;\;\;\;\;\;\;\;\;\;\;\;\;\;\;  \text{: unsafe set with margin $\gamma_\text{unsafe}>0$}  \label{eq:cons_2}\\
    &~~~~~~ \text{Lip}(h(z^i),\bar{\epsilon}) \leq L_h, \: \forall z^i \in Z_N \;\;\;\;\;\;\;\;\;\;\;\;\;\;\;\;\;\;\;\;\;\;\;\;\;\;\;\;\;\;\;\;\;\;\;\;\;\;  \text{: Lipschitz constraint on $h(z)$}  \label{eq:lip1}\\
    &~~~~~~  q_c(z^i,u^i_c):=\langle\nabla h(z^i), f_c(z^i)+g_c(z^i) u^i_c) \rangle \nonumber\\
    &~~~~~~~~~~~~~~~~~~~~~~~~~~~~~~~~~ \alpha(h(z^i)) \geq \gamma_\text{dyn}^c \;\;\;\; \text{: grad. constraint \eqref{eq:cbf_const_hybrid} with margin $\gamma_\text{dyn}^c>0$}\label{eq:cons_3}\\
    &~~~~~~ \text{Lip}(q_c(z^i,u^i_c),\epsilon_c) \leq L_q^c, \: \forall (z^i, u^i_c) \in Z_\text{dyn}^c \;\;\;\;\;\;\;\;\;\;\;\;\;\;\;\; \text{: Lipschitz constraint on $q_c(z^i,u^i_c)$} \label{eq:lip2_c}\\
     &~~~~~~  q_d(z^i,u^i_d):= h( f_d(z^i)+g_d(z^i) u^i_d) \geq \gamma_\text{dyn}^d \;\;\; \text{: jump constraint \eqref{eq:cbf_disc_hybrid} with margin $\gamma_\text{dyn}^d>0$} \label{eq:cons_4}\\
    &~~~~~~ \text{Lip}(q_d(z^i,u^i_d),\epsilon_d) \leq L_q^d, \: \forall (z^i, u^i_d) \in Z_\text{dyn}^d  \;\;\;\;\;\;\;\;\;\;\;\;\;\;\, \text{: Lipschitz constraint on $q_d(z^i,u^i_d)$}\label{eq:lip2_d}
\end{flalign}
\end{subequations}
where $\mathcal{H}$ is a normed function space and where the positive constants $\gamma_\text{safe}$, $\gamma_\text{unsafe}$, $\gamma_\text{dyn}^c$,$\gamma_\text{dyn}^d$, $L_h$, $L_q^c$, and $L_q^d$  are
\emph{hyperparameters} determined by the data-sets $Z_\text{safe}^c$, $Z_\text{safe}^d$, and $Z_N$, which must be sufficiently dense, as quantified by $\bar{\epsilon}$, $\epsilon_c$, and $\epsilon_d$ (conditions given below). 

\textbf{Lipschitz bounds:} The constraints in \eqref{eq:lip1}, \eqref{eq:lip2_c}, and \eqref{eq:lip2_d} assume a function $\text{Lip}(\cdot,\epsilon)$ that returns an upper bound on the Lipschitz constant of its argument with respect to the state $z$ in an $\epsilon$ neighborhood of $z^i$ using the $p$-norm as detailed in Appendix A. It may be difficult to enforce the constraints \eqref{eq:lip1}, \eqref{eq:lip2_c}, and \eqref{eq:lip2_d} in the optimization problem \eqref{eq:opt}. One can resort to bootstrapping the values of $L_h$, $L_q^c$, and $L_q^d$ by iteratively solving the optimization problem \eqref{eq:opt}, calculating the values of $L_h$, $L_q^c$, and $L_q^d$ to verify if constraints \eqref{eq:lip1}, \eqref{eq:lip2_c}, and \eqref{eq:lip2_d} hold, and adjusting regularization hyperparameters accordingly.  Appendix A explains how to compute Lipschitz constants for DNNs~\cite{fazlyab2019efficient} and RKHS. 

\textbf{Should we trust the experts?} Optimization problem \eqref{eq:opt} approximates the supremums over continuous and discrete inputs in the constraints \eqref{eq:cbf_const_hybrid} and \eqref{eq:cbf_disc_hybrid} with the expert actions -- while computationally expedient, this may prove conservative.  We note that in many cases of interest, the supremum over the continuous input $u_c \in \mathcal U_c$ in constraint \eqref{eq:cbf_const_hybrid} admits a closed form expression.  For example, when $\mathcal U_c$ is an $\|\cdot\|$-norm ball, the left hand side of constraint \eqref{eq:cbf_const_hybrid} reduces to  $\langle \nabla h(z^i), f_c(z^i) \rangle + \|\nabla h(z^i)g_c^T(z^i)\|_\star  + \alpha(h(z^i))$, for $\|\cdot\|_\star$ the dual norm.  This in turn can be used to simplify constraint \eqref{eq:cons_3} in optimization problem \eqref{eq:opt}, and in particular eliminates the dependency on $u^i_c$.  Nevertheless, the availability of expert demonstrations is still valuable as they indicate that a safe action exists, and we therefore expect a feasible HCBF $h$ and control action $u_c$ to exist.  
  
\textbf{Unconstrained relaxation:} For general function classes $\mathcal{H} := \{ h(z; \theta) | \theta \in \Theta \}$, e.g., when $h(z)$ is a deep neural net, optimization problem \eqref{eq:opt} is nonconvex: we therefore propose an unconstrained
relaxation that can be solved efficiently using stochastic first-order gradient methods
such as Adam or stochastic gradient descent. In particular, let $[r]_+ := \max \{r, 0\}$ for $r\in\mathbb{R}$ and relax the constrained optimization problem \eqref{eq:opt} to the unconstrained optimization problem
\begin{align*}
    &\min_{\theta \in \Theta} ~~ \|\theta\|^2 + \lambda_{\mathrm{s}} \sum_{z^i \in Z_\text{safe}^c\cup Z_\text{safe}^d} [\gamma_\text{safe} - h_\theta(z^i)]_+ 
    + \lambda_{\mathrm{u}} \sum_{z^i \in Z_N} [h_\theta(z^i) + \gamma_\text{unsafe}]_+ +\lambda_{\mathrm{d}} \sum_{(z^i, u^i_c) \in Z_\text{dyn}^c} [\gamma_\text{dyn}^c  \\
    & -  \langle \nabla h_\theta(z^i), f_c(z^i)+g_c(z^i)u_c^i \rangle - \alpha(h_\theta(z^i)) ]_+ + \lambda_{\mathrm{c}} \sum_{(z^i, u^i_d) \in Z_\text{dyn}^d} [\gamma_\text{dyn}^d -  h_\theta(f_d(z^i)+g_d(z^i)u_d^i) ]_+  
\end{align*}
where $\lambda_{\mathrm{s}}, \lambda_{\mathrm{u}},\lambda_{\mathrm{d}},\lambda_{\mathrm{c}}>0$ are hyperparameters that tradeoff between the constraints of problem \eqref{eq:opt}.


\paragraph{Guaranteeing Safety} We next show correctness of the learned HCBF $h(z)$  obtained from \eqref{eq:opt} in two steps by: 
\begin{enumerate}
    \item showing that the certified safe set \eqref{eq:local_C} is contained within the geometric safe set, i.e., that $\mathcal{C}\subset \mathcal{D}\subseteq \mathcal{S}$, and
    \item proving that $h(z)$ is a valid local HCBF by ensuring that the set $\mathcal{C}$ is forward invariant under control laws $u_c(z)\in\mathcal{U}_c$ and $u_d(z)\in\mathcal{U}_d$.
\end{enumerate}

\underline{1) Guaranteeing $\mathcal{C}\subset \mathcal{D}\subseteq \mathcal{S}$: } 
First assume that $Z_N $ is an $\bar{\epsilon}$-net of $\mathcal{N}$, i.e., for all $z \in \mathcal{N}$,  there exists $z^i\in Z_N$ such that $\|z^i-z\|_p\leq \bar{\epsilon}$. Using a standard covering and Lipschitz argument, we next show that if $h(z)$ satisfies constraint \eqref{eq:cons_2} for all $z^i\in Z_N$, we have that $h(z)<0$ for all $z\in\mathcal{N}$. 

\begin{proposition}\label{prop:1}
Let $h(z)$ be Lipschitz continuous with local constant $L_h(z)$\footnote{By local constant, we here mean a Lipschitz constant in an $\bar{\epsilon}$ neighborhood of $z$.}, $\gamma_\text{unsafe}>0$ and $Z_N $ be an $\bar{\epsilon}$-net of $\mathcal{N}$ with 
\begin{align*}
    \bar{\epsilon}<\gamma_\text{unsafe}/L_h(z^i) \ \text{ for all } \ z^i \in Z_N.
\end{align*} 
Then, if $h(z)$ satisfies constraint \eqref{eq:cons_2}, we have that   $h(z)<0$ for all $z\in\mathcal{N}$.
\begin{proof}
 Note first that, for all $z\in\mathcal{N}$, it follows that there exists a point $z^i \in Z_N$ satisfying $\|z-z^i\|_p\leq \bar{\epsilon}$ due to the assumption that $Z_N $ is an $\bar{\epsilon}$-net of $\mathcal{N}$. For any $z\in\mathcal{N}$, we can now select a point $z^i \in Z_N$ satisfying $\|z-z^i\|_p\leq \bar{\epsilon}$ for which it follows that
\begin{align*}
    h(z) &=  h(z)-h(z^i) + h(z^i) \overset{(a)}{\leq} |h(z)-h(z^i)| - \gamma_\text{unsafe}   \\&\overset{(b)}{\leq} L_h(z^i)\|z-z^i\|_p - \gamma_\text{unsafe} \overset{(c)}{\leq} L_h(z^i)\bar{\epsilon} - \gamma_\text{unsafe} \overset{(d)}{<} 0
\end{align*}
In particular, inequality $(a)$ follows from the constraint \eqref{eq:cons_2} which says that $h(z^i)\leq-\gamma_\text{unsafe}$ for all $z^i\in Z_N$. Inequality $(b)$ follows by the local Lipschitz constant $L_h(z)$ on $h(z)$ in an $\bar{\epsilon}$ neighborhood of $z^i$, while inequality $(c)$ follows again by the assumption that $Z_N$ forms an $\bar{\epsilon}$-net of $\mathcal{N}$. The strict inequality $(d)$ follows simply by the assumption that $\bar{\epsilon}<\gamma_\text{unsafe}/L_h(z^i)$ for all $z^i \in Z_N$.
\end{proof}
\end{proposition}

We can use a similar argument on constraint \eqref{eq:cons_1}, which ensures that the set $\mathcal{C}$ over which $h(z)\geq 0$, as defined in equation \eqref{eq:local_C}, has non-empty interior.  
\begin{proposition}\label{cor:1}
Let $h(z)$ be Lipschitz continuous with local constant $L_h(z)$, and $Z_\text{safe}^c$ and $Z_\text{safe}^d$ be ${\epsilon}_c$- and ${\epsilon}_d$-nets of $\mathcal{D}_C$ and $\mathcal{D}_D$, respectively, with 
\begin{align*}
    \max({\epsilon}_c,{\epsilon}_d)\leq\gamma_\text{safe}/L_h(z^i) \ \text{ for all } \ z^i \in Z_\text{safe}^c\cup Z_\text{safe}^d
\end{align*} 
and $\gamma_\text{safe}>0$. Then, if $h(z)$ satisfies constraint \eqref{eq:cons_1}, we have that $h(z)\geq 0$ for all $z\in\mathcal{D}$.

\begin{proof}
 Note first that, for all $z\in \mathcal{D}_C$, it again follows that there exists a point $z^i \in Z_\text{safe}^c$ satisfying $\|z-z^i\|_p\leq \epsilon_c$ due to the assumption that $Z_\text{safe}^c$ is an $\epsilon_c$-net of $\mathcal{D}_C$. For any $z\in\mathcal{D}_C$, we can now select a point $z^i \in Z_\text{safe}^c$ satisfying $\|z-z^i\|_p\leq \epsilon_c$ for which it follows that
\begin{align*}
    0&\overset{(a)}{\leq} h(z^i)-\gamma_\text{safe}=h(z^i)-h(z)+h(z)-\gamma_\text{safe} \leq|h(z^i)-h(z)|+h(z)-\gamma_\text{safe}\\
    &\overset{(b)}{\leq}L_h(z^i)\|z^i-z\|_p+h(z)-\gamma_\text{safe}\overset{(c)}{\leq}L_h(z^i)\epsilon_c+h(z)-\gamma_\text{safe}\overset{(d)}{\leq}h(z)
\end{align*}
In particular, inequality $(a)$ follows from the constraint \eqref{eq:cons_1} which says that $h(z^i)\ge\gamma_\text{safe}$ for all $z^i\in Z_\text{safe}^c$. Inequality $(b)$ follows by the local Lipschitz constant $L_h(z)$ on $h(z)$ in an $\epsilon_c$ neighborhood of $z^i$, while inequality $(c)$ follows again by the assumption that $Z_\text{safe}^c$ is an $\epsilon_c$-net of $\mathcal{D}_C$. The inequality $(d)$ follows simply by the assumption that $\max({\epsilon}_c,{\epsilon}_d)\leq\gamma_\text{safe}/L_h(z^i)$ for all $z^i \in Z_\text{safe}^c\cup Z_\text{safe}^d$. The same analysis holds for all $z\in \mathcal{D}_D$, so that $h(z)\geq 0$ for all $z\in\mathcal{D}$.
\end{proof}
\end{proposition}

Propositions \ref{prop:1} and \ref{cor:1} then ensure that $\mathcal{C}\subset\mathcal{D}\subseteq\mathcal{S}$, i.e., the zero level-set of $h(z)$ is contained within the geometric safe set $\mathcal S$.  Note that the constraints \eqref{eq:cons_1} and \eqref{eq:cons_2} may, in practice, lead to infeasibility of \eqref{eq:opt}. We resort to the same remedy as proposed in \cite[equation (3.4)]{robey2020learning}, i.e., enforcing constraint \eqref{eq:cons_1} on smaller sets $\bar{Z}_\text{safe}^c$ and $\bar{Z}_\text{safe}^d$ with $\bar{Z}_\text{safe}^c\subset Z_\text{safe}^c$ and $\bar{Z}_\text{safe}^d\subset Z_\text{safe}^d$ to allow for smoother HCBFs to be learned at the expense of a smaller invariant safe set $\mathcal{C}$, i.e., replace \eqref{eq:cons_1} by
\begin{align}\label{eq:cons_1_bar}
h(z^i) \geq \gamma_\text{safe}, \: \forall z^i \in \bar{Z}_\text{safe}^c\cup \bar{Z}_\text{safe}^d.
\end{align}

\underline{2) Guaranteeing a valid local HCBF: } For a state $z\in \mathcal{D}_C$, recall the definition of the function 
\begin{align*}
    q_c(z,u_c^i):= \langle \nabla h(z),f_c(z)+g_c(z)u^i_c\rangle+\alpha(h(z))
\end{align*} 
where the control sample $u_c^i$ is associated with the sample $z^i$ that is such that $\|z^i-z\|_p\le \epsilon_c$. Note that such a pair $(z^i,u_c^i)\in Z_\text{dyn}^c$ is guaranteed to exist when the set $Z_\text{safe}^c$ is an $\epsilon_c$-net of $\mathcal{D}_D$. The function $q_c(z,u_c^i)$ is Lipschitz continuous in $z$ with local constant denoted by $L_q^c(z,u_c^i)$, as we have assumed $h$ to be twice continuously differentiable. Recall also that 
\begin{align*}
    q_d(z,u_d^i):= h(f_d(z)+g_d(z)u^i_d)
\end{align*}
    and note similarly that $q_d(z,u_d^i)$ is Lipschitz continuous with local constant denoted by $L_q^d(z,u_d^i)$. We next provide conditions guaranteeing that the learned HCBF satisfies the constraint \eqref{eq:cbf_const_hybrid} for all $z \in \mathcal{D}_C$ and the constraint \eqref{eq:cbf_disc_hybrid} for all $z \in \mathcal{D}_D$.

\begin{proposition}\label{prop:2}
Suppose $q_c(z,u_c^i)$ and $q_d(z,u_d^i)$ are Lipschitz continuous with local constants $L_q^c(z,u_c^i)$ and $L_q^d(z,u_d^i)$, respectively.  Let $\gamma_\text{dyn}^c,\gamma_\text{dyn}^d>0$, and assume that (i) $Z_\text{safe}^c$ is an $\epsilon_c$-net of $\mathcal{D}_C$ with 
\begin{align*}
   \epsilon_c\leq \gamma_\text{dyn}^c/L_q^c(z^i,u_c^i) \ \text{ for all } \ (z^i,u_c^i)\in Z_\text{dyn}^c, 
\end{align*} 
and (ii) $Z_\text{safe}^d$ is an $\epsilon_d$-net of $\mathcal{D}_D$ with 
\begin{align*}
    \epsilon_d\leq \gamma_\text{dyn}^d/L_q^d(z^i,u_d^i) \ \text{ for all } \ (z^i,u_d^i)\in Z_\text{dyn}^d.
\end{align*}
Then, if $h(z)$ satisfies constraints \eqref{eq:cons_3} and \eqref{eq:cons_4}, we have that $q_c(z,u_c^i)\geq 0$ for all $z \in \mathcal{D}_C$ and $q_d(z,u_d^i)\geq 0$ for all $z \in \mathcal{D}_D$.

\begin{proof}
 Note first that, for all $z\in\mathcal{D}_C$, it follows that there exists a pair $(z^i,u_c^i) \in Z_\text{dyn}^c$ satisfying $\|z-z^i\|_p\leq \epsilon_c$ due to the assumption that $Z_\text{safe}^c$ is an $\epsilon_c$-net of $\mathcal{D}_C$. For any $z\in\mathcal{D}_C$, we can now select a pair $(z^i,u_c^i) \in Z_\text{dyn}^c$ satisfying $\|z-z^i\|_p\leq \epsilon_c$ for which it follows that
\begin{align*}
    0&\overset{(a)}{\leq} q_c(z^i,u_c^i)-\gamma_\text{dyn}^c=q_c(z^i,u_c^i)-q_c(z,u_c^i)+q_c(z,u_c^i)-\gamma_\text{dyn}^c \\&
    \leq|q_c(z^i,u_c^i)-q_c(z,u_c^i)|+q_c(z,u_c^i)-\gamma_\text{dyn}^c\overset{(b)}{\leq}L_q^c(z^i,u_c^i)\|z^i-z\|_p+q_c(z,u_c^i)-\gamma_\text{dyn}^c\\
    &\overset{(c)}{\leq}L_q^c(z^i,u_c^i)\epsilon_c+q_c(z,u_c^i)-\gamma_\text{dyn}^c\overset{(d)}{\leq}q_c(z,u_c^i)
\end{align*}
In particular, inequality $(a)$ follows from the constraint \eqref{eq:cons_3} which says that $q_c(z^i,u_c^i)\ge\gamma_\text{dyn}^c$ for all $(z^i,u_c^i) \in Z_\text{dyn}^c$. Inequality $(b)$ follows by the local Lipschitz constant $L_q^c(z,u_c^i)$ on $q_c(z,u_c^i)$ in an $\epsilon_c$ neighborhood of $z^i$, while inequality $(c)$ follows again by the assumption that $Z_\text{safe}^c$ is an $\epsilon_c$-net of $\mathcal{D}_C$. The inequality $(d)$ follows simply by the assumption that $\epsilon_c\leq\gamma_\text{dyn}^c/L_q^c(z^i,u_c^i)$ for all $z^i \in Z_\text{safe}^c$. This implies that $q_c(z,u_c^i)\ge 0$ for all $z\in\mathcal{D}_C$. The same analysis holds for all $z\in \mathcal{D}_D$, so that $q_d(z,u_d^i)\ge 0$ for all $z\in\mathcal{D}$.
\end{proof}
\end{proposition}

By combining the previous arguments, we can ensure that the learned function $h(z)$ defines an appropriate safe set, as captured by the condition $\mathcal{C} \subset \mathcal{D} \subseteq \mathcal{S}$, that can be rendered forward invariant by choosing control actions satisfying the constraints \eqref{eq:cbf_const_hybrid} and \eqref{eq:cbf_disc_hybrid}. The next theorem summarizes these results and guarantees that $h(z)$ from \eqref{eq:opt} is a valid HCBF.

\begin{theorem}\label{thm:2}
Let $h(z)$ be a twice continuously differentiable function and let the sets $\mathcal{S}$, $\mathcal{N}$, $\mathcal{D}_C$, $\mathcal{D}_D$, $\mathcal{D}$, $\mathcal{C}$, and the data-sets $Z_\text{dyn}^c$, $Z_\text{dyn}^d$, $Z_\text{safe}^c$, $Z_\text{safe}^d$, and $Z_N$ be defined as above. Suppose that $Z_N$ forms an $\bar{\epsilon}$-net of $\mathcal{N}$ satisfying $\bar{\epsilon}<\gamma_\text{unsafe}/L_h(z^i)$ for all $z^i \in Z_N$, and that $Z_\text{safe}^c$ and $Z_\text{safe}^d$ are ${\epsilon}_c$- and  ${\epsilon}_d$-nets of $\mathcal{D}_C$ and $\mathcal{D}_D$, respectively, satisfying the conditions of Propositions \ref{cor:1} \& \ref{prop:2}. Let $h(z)$, $q_c(z,u_c^i)$, and $q_d(z,u_c^i)$ be Lipschitz continuous with local constants $L_h(z)$, $L_q^c(z,u_c^i)$, and $L_q^d(z,u_d^i)$, respectively. Then if $h(z)$ satisfies constraints \eqref{eq:cons_1_bar}, \eqref{eq:cons_2}, \eqref{eq:cons_3}, and \eqref{eq:cons_4}, the set $\mathcal{C}$ is non-empty, $\mathcal{C}\subset\mathcal{D}\subseteq\mathcal{S}$, and the function $h(z)$ is a valid local hybrid control barrier function on $\mathcal{D}$ with domain $\mathcal{N} \cup \mathcal{D}$.
\end{theorem}


\section{Case Studies}
\label{sec:case_studies}

The code for both case studies is available at \href{https://github.com/unstable-zeros/learning-hcbfs}{https://github.com/unstable-zeros/learning-hcbfs}.
In both our simulations, we numerically
integrate the hybrid dynamics of interest using an integrator implementation inspired by
Drake~\cite{drake}. We build our implementation on top of
\texttt{scipy.integrate.solve\_ivp}'s 
event detection API. This allows us to 
get precise notifications for when a system makes
a discrete jump.

\begin{figure}[h!]
\centering
\includegraphics[scale=0.53]{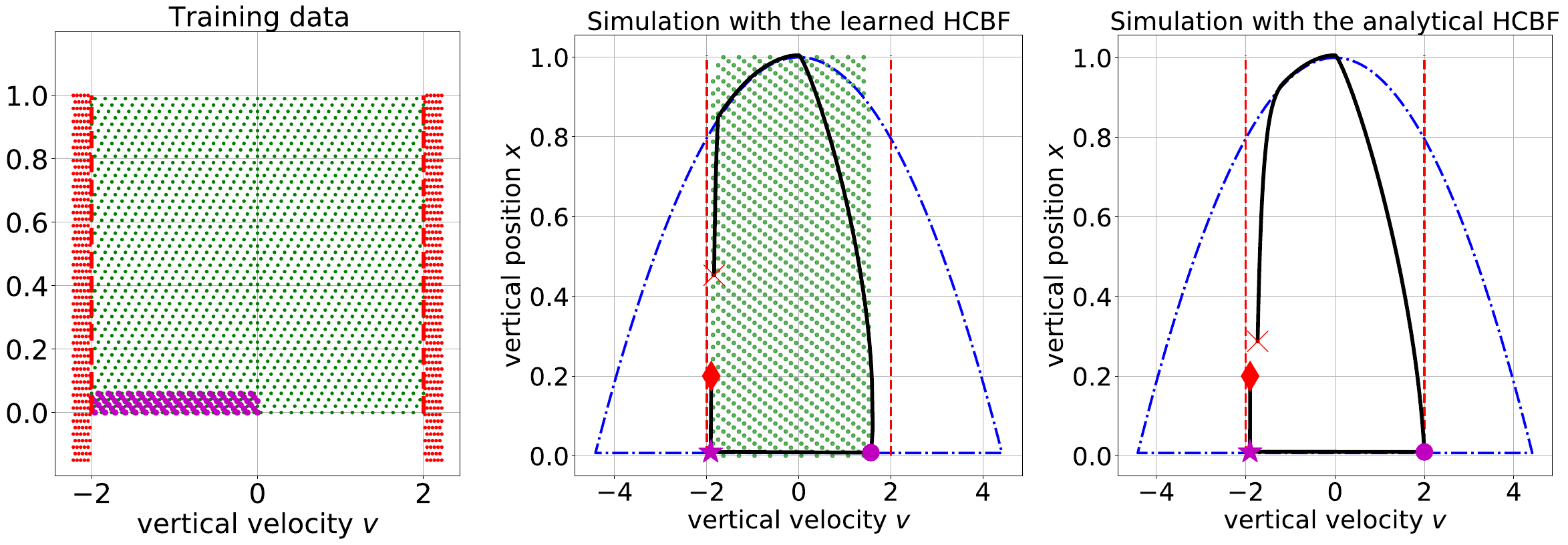}
\caption{Left: Plot of the expert states $Z^c_{\text{safe}}$ (green) and $Z^d_{\text{safe}}$ (magenta), and unsafe samples $Z_N$ (red). Dashed red lines indicate bounds on the velocity.
Middle: Closed-loop simulation of the bouncing ball using the learned HCBF where the initial state, state before jump, state after jump and terminal state are indicated by a red diamond, magenta star, magenta circle and red cross, respectively. The solid black line and dash-dotted blue line represent the closed-loop trajectory and desired reference path. The green dots denote the safe invariant set \eqref{eq:local_C} of the learned HCBF.
Right: Closed-loop simulation of the bouncing ball using the analytical HCBF.}
\label{fig:ball}
\end{figure}

\paragraph{Bouncing ball}
Our first experiment considers the controlled bouncing ball discussed in Example~\ref{ex:2}.
Here we define a safe set $\mathcal{S}:=\{z\in\mathbb{R}^2| | v | \le 2\}$, and set $\rho=9.81$.
Details as to how expert data is generated can be found in Appendix B. At a high level, we use an exploratory controller that covers much of the state-space but does not guarantee safety, in combination with a safe controller composed of a continuous component based on the analytical HCBF in Example \ref{ex:2} and the CBF-QP problem \cite{ames2014control}, and a discrete component obtained by analytically solving the constraint described in Example \ref{ex:2}.
Using the above controllers, we obtain data-sets $Z^c_{\text{safe}}$ and $Z^d_{\text{safe}}$ containing 5000 safe states and expert demonstrations $u^i_c$ and $u^i_d$.
Those expert states, shown as the green and magenta dots in Fig.~\ref{fig:ball}(left), are evenly gridded near the boundary of the safe set $\mathcal{S}$.
We in addition sample 4560 unsafe samples by gridding along the boundary of $\mathcal{N}$ to form the data-set $Z_N$.

We parametrize the HCBF candidate $h$ as a two-hidden-layer fully-connected DNN with tanh activation functions and 64 neurons in each hidden layer. The training is implemented using \texttt{jax}~\cite{bradbury2020jax} and the Adam algorithm with a cosine decay learning rate. 
We craft a loss function by relaxing the constraints in \eqref{eq:opt} -- hyperparameter choices and other training details can be found in Appendix B.
The set $\mathcal{C}$ of the learned HCBF is shown as green dots in the middle subplot of Figure \ref{fig:ball}.
One may clearly observe that $\mathcal{C}$ is strictly contained within the geometric safe set, i.e. $\mathcal{C} \subset \mathcal{S}$.

We now test if the learned HCBF is able to produce safe control inputs that keep the system within the set $\mathcal{C}$.
We consider a nominal control law $u_{c,\text{nom}}$ that looks to track the reference path shown in dash-dotted blue in Fig.~\ref{fig:ball}(middle) and (right): as the system has full control, it can exactly track the reference path, but without correction, this would lead to violation of the velocity constraint.
Starting from an initial state $(0.2, -1.9)$, we simulate the controlled system for $2.25$ seconds.
During flow, we solve the CBF-QP problem for the continuous input $u_{c}$ (see Appendix B.2).
During jump, we perform line search with a decay factor $0.95$ to find the scalar discrete input $u_{d}$ that satisfies \eqref{eq:cbf_disc_hybrid}.
The closed-loop trajectory produced by the learned HCBF is plotted in Figure \ref{fig:ball} (middle).
As a comparison, we also plot in Figure \ref{fig:ball} (right) the trajectory obtained using the analytical HCBF.
Albeit slightly more conservative, the learned HCBF keeps the ball within $\mathcal{C}$ at all times during both flow and jump.

\paragraph{Compass gait}

We consider the compass gait walker \cite{goswami1996limit,byl2008approximate}.  Originally introduced by~\cite{goswami1996limit}, the compass gait dynamics describe a passive bipedal robot walking down an inclined plane at a constant velocity.  This system is described by a four-dimensional continuous state 
\begin{align*}
    z = [\theta_{\text{stance}}, \theta_{\text{swing}}, \dot{\theta}_{\text{stance}}, \dot{\theta}_{\text{swing}}],
\end{align*}
consisting of the angle $\theta$ and angular velocity $\dot{\theta}$ of each leg.  In this notation, the ``stance'' foot corresponds to the foot that is in contact with the ground as the compass gait walker makes its descent down the ramp; hence, the ``swing'' foot refers to the foot that is not in contact with the ramp at a particular instant in time.  In our simulations, the walker's initial stance leg is its left leg, and therefore its initial swing leg is its right leg.  To improve the walking capabilities, we add actuation to hip joint and the ankle of the stance leg.  To collect expert trajectories, we use the energy-based controller of~\cite{goswami1997limit}.   All implementation details for the compass gait walker can be found in Appendix C.  

Learning and analyzing an HCBF for the compass gait walker hybrid system poses several challenges, including the well-known sensitivity of the compass gait walker to its initial conditions and the inherent difficulties in visualizing a four-dimensional state space.  To facilitate a meaningful visualization of the four-dimensional state space, when collecting expert data, we fix the stance leg initial condition to the point $[\theta_{\text{stance}}, \dot{\theta}_{\text{stance}}] = [0, 0.4]$ on the passive limit cycle, and vary the initial condition of the swing leg by adding uniform noise to corresponding passive limit cycle state $[\theta_{\text{swing}}, \dot{\theta}_{\text{swing}}] = [0, 2.0]$ of the swing leg.

\begin{figure}
    \centering
    \includegraphics[width=\textwidth]{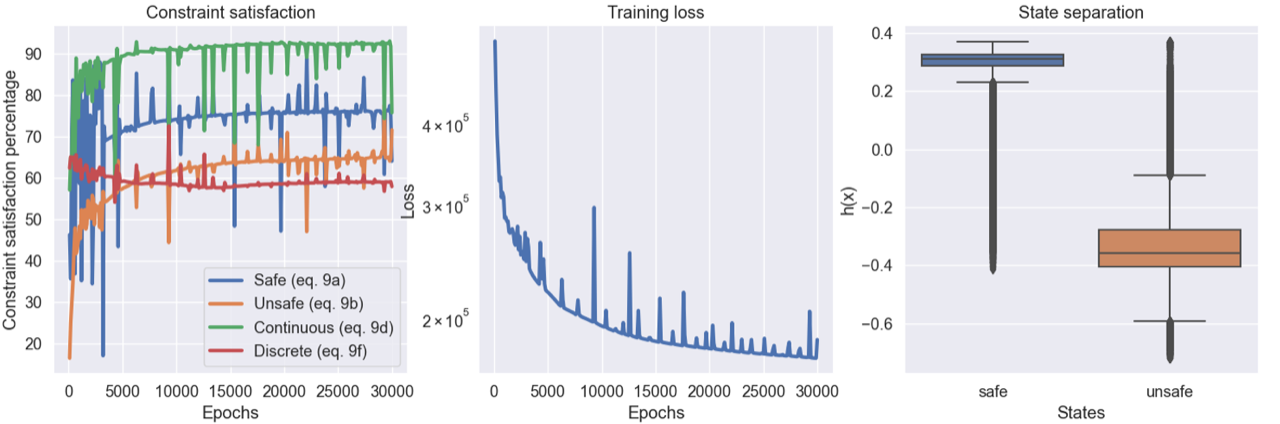}
    \caption{Left: We show the constraint satisfaction rates for the soft constraints imposed in the unconstrained relaxation of \eqref{eq:opt} as described in the paper while training the HCBF.  Middle: We plot the training loss as a function of the training epoch.  Right: We show boxplots corresponding to the values of the trained HCBF on the safe and unsafe states.  Notice that for states marked safe, we generally have $h(z) > 0$, whereas for states marked unsafe, we generally have $h(z) < 0$.}
    \label{fig:train-compass-gait}
\end{figure}

We again parameterize the candidate HCBF $h$ with a two-hidden-layer fully-connected neural network with tanh activations and with 32 and 16 neurons in the first and second hidden layers, respectively; as before, we determine the hyperpameters by grid search.  However, unlike the previous example, the task of identifying boundary points is complicated by the higher dimensionality of the state space.  Therefore, we propose a novel algorithm that can be used to identify boundary points by sampling from the space of expert trajectories.  The algorithm, described in Appendix C, identifies boundary points by computing the pairwise distances between all of the expert states and thresholding based on the number of neighbors a point has within an $\epsilon$-norm ball.

In Figure \ref{fig:train-compass-gait}, we show the constraint satisfaction rates, the training loss, and the state separation attained by the learned HCBF. Note that these figures highlight the challenges of multi-objective optimization, i.e., trying to achieve all of the constraints in the optimization problem \eqref{eq:opt} via its unconstrained relaxation. Nevertheless, we emphasize that the use of robust optimization by including slack variables $\gamma_\text{safe}$, $\gamma_\text{unsafe}$, $\gamma_\text{dyn}^c$, and $\gamma_\text{dyn}^d$ leads to HCBFs that perform well in practice.

\begin{figure}
    \centering
    \begin{subfigure}{0.4\textwidth}
        \includegraphics[width=\textwidth]{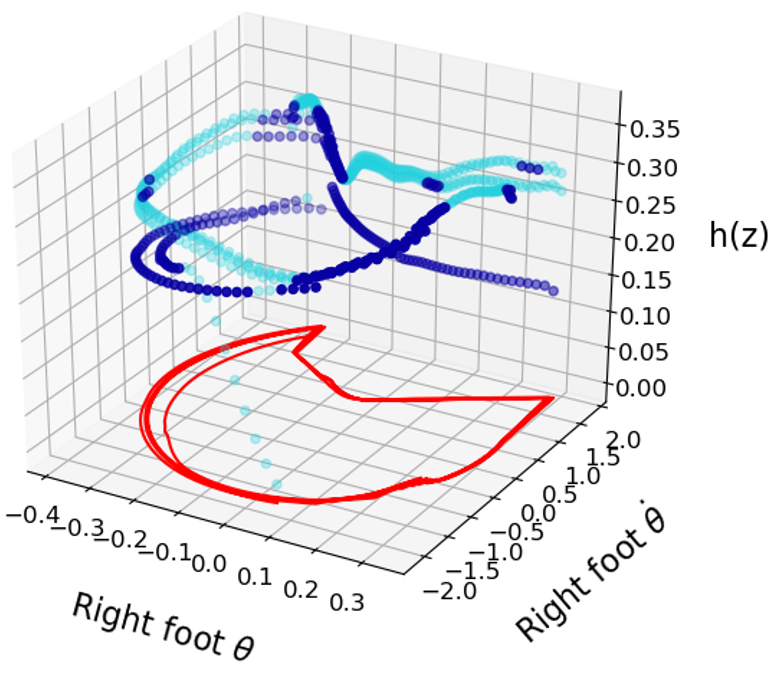}
    \end{subfigure} \quad\quad
    \begin{subfigure}{0.45\textwidth}
        \includegraphics[width=\textwidth]{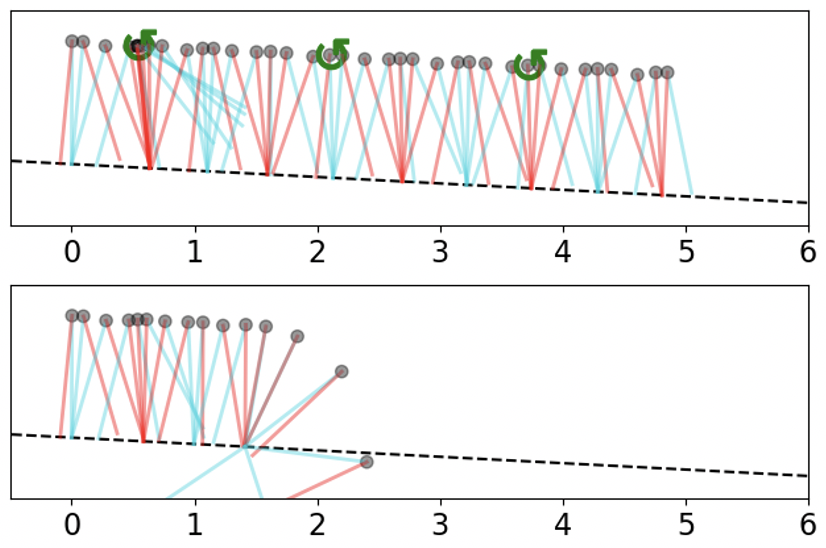}
    \end{subfigure}
    \caption{(Left) In red, we show the phase portrait of the right foot using the HCBF-QP controller.  On the vertical axis, we plot the values of $h(z)$ associated with this phase portrait.  The regions where $u_{\text{HCBF}} \neq u_{\text{nom}}$, i.e., where the HCBF corrects the nominal controller, are shown in dark blue; values of $h(x)$ where $u_{\text{HCBF}} = u_{\text{nom}}$ are shown in light blue.  Notice that when the value of $h(z)$ dips below the ``safe'' threshold $\gamma_{\text{safe}} = 0.3$, the HCBF-QP almost always intervenes.  (Right) From a fixed initial state, we show a trajectory with the HCBF-QP controller (top) and the nominal zero-valued controller (bottom).  The states are down-sampled for clarity, the left and right feet are shown in red and blue respectively, and torques applied by the HCBF-QP controller are shown as green arrows.}
    \label{fig:compass-gait}
\end{figure}

In the left panel of Fig.~\ref{fig:compass-gait}, we visualize the learned HCBF.  Starting from an initial condition with the same left leg state as the expert trajectories, we identify the values for $h(z)$ at which $u_{\text{HCBF}} \neq u_{\text{nom}}$ in green, i.e. where the HCBF controller corrects the nominal controller. In red, we plot the phase portrait for the right leg corresponding to a trajectory using the HCBF-QP controller.  To demonstrate the physical interpretation of this learned HCBF, in the right top panel of Fig.~\ref{fig:compass-gait}, we show the motion of the compass gait walker down the ramp, marking in green where the HCBF-based controller takes over; in the right lower panel of Fig.~\ref{fig:compass-gait} we see the failure of a zero-valued controller from the same initial condition, showing that the learned HCBF preserves safety.  Videos of both the safe HCBF and unsafe nominal controller trajectories can be found in the supplementary material.

\begin{figure}
    \centering
    \includegraphics[width=0.9\textwidth]{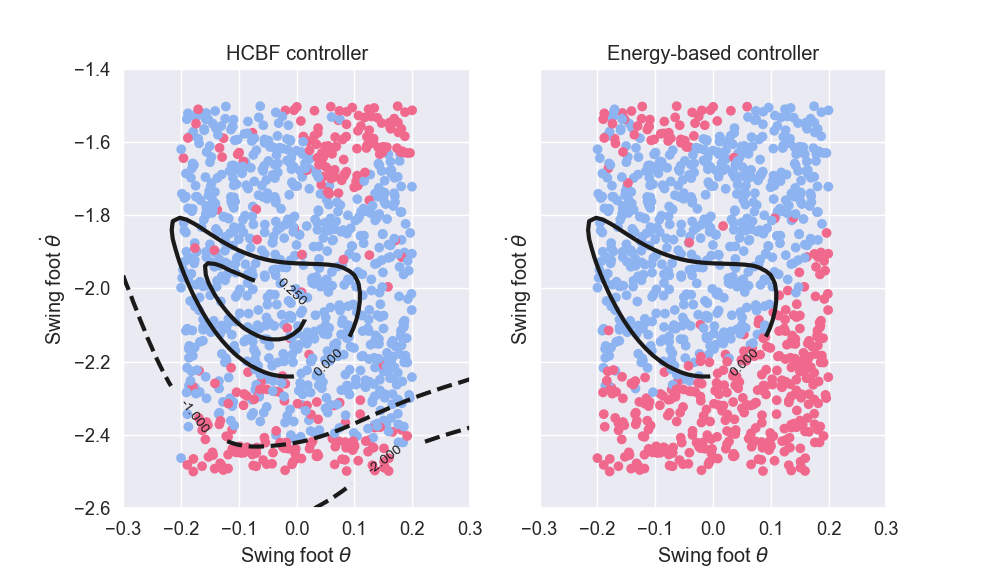}
    \caption{For a fixed standing foot initial condition of $[\theta_{\text{stance}}, \dot{\theta}_{\text{stance}}] = [0, 0.4]$, we vary the swing foot initial condition and simulate the compass gait walker with the HCBF controller on top of a nominal zero-valued controller (left) and the energy-based controller (right).  Each blue dot denotes a swing foot initial condition that corresponds to a trajectory in which the walker travels eight or more steps; red dots denote trajectories in which the walker travels fewer than eight steps.  We also plot the level sets of the learned HCBF for reference.}
    \label{fig:init-conds-compass-gait}
\end{figure}

To visualize the safe set of the learned HCBF, in Figure \ref{fig:init-conds-compass-gait}, we show that despite the fact that our nominal controller is not providing any control inputs, the HCBF-based controller has a similar safe set to that of the energy-based controller.  In this way, the HCBF causes the uncontrolled system to match the safety characteristics of the expert energy-based controller.  In particular, it can be observed that the safe set $\mathcal{C}$, which is described by the zero superlevel set of $h(z)$,  under-approximates the safe expert behavior. As can further be observed, there are regions in the state space where our HCBF-based controller provides safe system trajectories while the energy-based controller results in unsafe system trajectories (e.g., the bottom right corner of both plots shown in Figure \ref{fig:init-conds-compass-gait}). We suspect that the safe set $\mathcal{C}$ enjoys asymptotic stability properties similar to those of the continuous case \cite{ames2017control}, allowing for an expanded region of safety as compared to that of the expert. A more detailed investigation of this observation is, however, subject to future work.

\section{Conclusion}
\label{sec:conclusion}

We presented a framework for learning safe control laws for hybrid systems. We introduced HCBFs and sufficient conditions under which an HCBF based control law guarantees safety, i.e., a desired safe set is made forward invariant. We then showed how to learn such HCBFs from data using an optimization-based framework. We gave sufficient conditions under which feasible solutions to the optimization problem are valid HCBFs, and illustrated our methods on two case studies.  
Future work will look to extend this approach to hybrid systems described by differential \emph{inclusions}, so as to be applicable to systems with friction and stiction, as well as to develop statistical guarantees of correctness to alleviate the sampling burden of our method.

\clearpage

\newpage
\bibliographystyle{IEEEtran}
\bibliography{literature}

\begin{thebibliography}{10}
\providecommand{\url}[1]{#1}
\csname url@samestyle\endcsname
\providecommand{\newblock}{\relax}
\providecommand{\bibinfo}[2]{#2}
\providecommand{\BIBentrySTDinterwordspacing}{\spaceskip=0pt\relax}
\providecommand{\BIBentryALTinterwordstretchfactor}{4}
\providecommand{\BIBentryALTinterwordspacing}{\spaceskip=\fontdimen2\font plus
\BIBentryALTinterwordstretchfactor\fontdimen3\font minus
  \fontdimen4\font\relax}
\providecommand{\BIBforeignlanguage}[2]{{%
\expandafter\ifx\csname l@#1\endcsname\relax
\typeout{** WARNING: IEEEtran.bst: No hyphenation pattern has been}%
\typeout{** loaded for the language `#1'. Using the pattern for}%
\typeout{** the default language instead.}%
\else
\language=\csname l@#1\endcsname
\fi
#2}}
\providecommand{\BIBdecl}{\relax}
\BIBdecl

\bibitem{schwarting2018planning}
W.~Schwarting, J.~Alonso-Mora, and D.~Rus, ``Planning and decision-making for
  autonomous vehicles,'' \emph{Annual Review of Control, Robotics, and
  Autonomous Systems}, 2018.

\bibitem{tucker2019preference}
M.~Tucker, E.~Novoseller, C.~Kann, Y.~Sui, Y.~Yue, J.~Burdick, and A.~D. Ames,
  ``Preference-based learning for exoskeleton gait optimization,'' \emph{arXiv
  preprint arXiv:1909.12316}, 2019.

\bibitem{kress2009temporal}
H.~Kress-Gazit, G.~E. Fainekos, and G.~J. Pappas, ``Temporal-logic-based
  reactive mission and motion planning,'' \emph{IEEE Trans. Robot.}, vol.~25,
  no.~6, pp. 1370--1381, 2009.

\bibitem{prajna2007framework}
S.~Prajna, A.~Jadbabaie, and G.~J. Pappas, ``A framework for worst-case and
  stochastic safety verification using barrier certificates,'' \emph{IEEE
  Trans. Autom. Control}, vol.~52, no.~8, pp. 1415--1428, 2007.

\bibitem{glotfelter2017nonsmooth}
P.~Glotfelter, J.~Cort{\'e}s, and M.~Egerstedt, ``Nonsmooth barrier functions
  with applications to multi-robot systems,'' \emph{IEEE Control Syst. Lett.},
  vol.~1, no.~2, pp. 310--315, 2017.

\bibitem{glotfelter2019hybrid}
P.~Glotfelter, I.~Buckley, and M.~Egerstedt, ``Hybrid nonsmooth barrier
  functions with applications to provably safe and composable collision
  avoidance for robotic systems,'' \emph{IEEE Robotics and Automation Letters},
  vol.~4, no.~2, pp. 1303--1310, 2019.

\bibitem{wieland2007constructive}
P.~Wieland and F.~Allg{\"o}wer, ``Constructive safety using control barrier
  functions,'' in \emph{Proc. IFAC Symp. Nonlin. Control Syst.}, Pretoria,
  South Africa, August 2007, pp. 462--467.

\bibitem{ames2014control}
A.~D. Ames, J.~W. Grizzle, and P.~Tabuada, ``Control barrier function based
  quadratic programs with application to adaptive cruise control,'' in
  \emph{Proceedings of the Conference on Decision and Control (CDC)}, Los
  Angeles, CA,, December 2014, pp. 6271--6278.

\bibitem{ames2017control}
A.~D. Ames, X.~Xu, J.~W. Grizzle, and P.~Tabuada, ``Control barrier function
  based quadratic programs for safety critical systems,'' \emph{IEEE Trans.
  Autom. Control}, vol.~62, no.~8, pp. 3861--3876, 2017.

\bibitem{khojasteh2019probabilistic}
M.~J. Khojasteh, V.~Dhiman, M.~Franceschetti, and N.~Atanasov, ``Probabilistic
  safety constraints for learned high relative degree system dynamics,''
  \emph{arXiv preprint arXiv:1912.10116}, 2019.

\bibitem{taylor2020control}
A.~J. Taylor, A.~Singletary, Y.~Yue, and A.~D. Ames, ``A control barrier
  perspective on episodic learning via projection-to-state safety,''
  \emph{arXiv preprint arXiv:2003.08028}, 2020.

\bibitem{agrawal2017discrete}
A.~Agrawal and K.~Sreenath, ``Discrete control barrier functions for
  safety-critical control of discrete systems with application to bipedal robot
  navigation.'' in \emph{Robotics: Science and Systems}, 2017.

\bibitem{ohnishi2019barrier}
M.~Ohnishi, L.~Wang, G.~Notomista, and M.~Egerstedt, ``Barrier-certified
  adaptive reinforcement learning with applications to brushbot navigation,''
  \emph{IEEE Transactions on robotics}, vol.~35, no.~5, pp. 1186--1205, 2019.

\bibitem{cavorsi2020tractable}
M.~Cavorsi, M.~Khajenejad, R.~Niu, Q.~Shen, and S.~Z. Yong, ``Tractable
  compositions of discrete-time control barrier functions with application to
  lane keeping and obstacle avoidance,'' \emph{arXiv preprint
  arXiv:2004.01858}, 2020.

\bibitem{maghenem2019characterizations}
M.~Maghenem and R.~G. Sanfelice, ``Characterizations of safety in hybrid
  inclusions via barrier functions,'' in \emph{Proceedings of the 22nd ACM
  International Conference on Hybrid Systems: Computation and Control}, 2019,
  pp. 109--118.

\bibitem{bisoffi2018hybrid}
A.~Bisoffi and D.~V. Dimarogonas, ``A hybrid barrier certificate approach to
  satisfy linear temporal logic specifications,'' in \emph{2018 Annual American
  Control Conference (ACC)}.\hskip 1em plus 0.5em minus 0.4em\relax IEEE, 2018,
  pp. 634--639.

\bibitem{goebel2012hybrid}
R.~Goebel, R.~G. Sanfelice, and A.~R. Teel, \emph{Hybrid Dynamical Systems:
  modeling, stability, and robustness}, 1st~ed.\hskip 1em plus 0.5em minus
  0.4em\relax Princeton, NJ: Princeton University Press, 2012.

\bibitem{lygeros2003dynamical}
J.~Lygeros, K.~H. Johansson, S.~N. Simic, J.~Zhang, and S.~S. Sastry,
  ``Dynamical properties of hybrid automata,'' \emph{IEEE Transactions on
  automatic control}, vol.~48, no.~1, pp. 2--17, 2003.

\bibitem{ames2019control}
A.~D. Ames, S.~Coogan, M.~Egerstedt, G.~Notomista, K.~Sreenath, and P.~Tabuada,
  ``Control barrier functions: Theory and applications,'' in \emph{2019 18th
  European Control Conference (ECC)}, Naples, Italy, June 2019, pp. 3420--3431.

\bibitem{xu2017correctness}
X.~Xu, J.~W. Grizzle, P.~Tabuada, and A.~D. Ames, ``Correctness guarantees for
  the composition of lane keeping and adaptive cruise control,'' \emph{IEEE
  Transactions on Automation Science and Engineering}, vol.~15, no.~3, pp.
  1216--1229, 2017.

\bibitem{wang2018permissive}
L.~Wang, D.~Han, and M.~Egerstedt, ``Permissive barrier certificates for safe
  stabilization using sum-of-squares,'' in \emph{2018 Annual American Control
  Conference (ACC)}.\hskip 1em plus 0.5em minus 0.4em\relax IEEE, 2018, pp.
  585--590.

\bibitem{yaghoubi2020training}
S.~Yaghoubi, G.~Fainekos, and S.~Sankaranarayanan, ``Training neural network
  controllers using control barrier functions in the presence of
  disturbances,'' \emph{arXiv preprint arXiv:2001.08088}, 2020.

\bibitem{srinivasan2020synthesis}
M.~Srinivasan, A.~Dabholkar, S.~Coogan, and P.~Vela, ``Synthesis of control
  barrier functions using a supervised machine learning approach,'' \emph{arXiv
  preprint arXiv:2003.04950}, 2020.

\bibitem{saveriano2019learning}
M.~Saveriano and D.~Lee, ``Learning barrier functions for constrained motion
  planning with dynamical systems,'' in \emph{IEEE International Conference on
  Intelligent Robots and Systems}, 2019.

\bibitem{jin2020neural}
W.~Jin, Z.~Wang, Z.~Yang, and S.~Mou, ``Neural certificates for safe control
  policies,'' \emph{arXiv preprint arXiv:2006.08465}, 2020.

\bibitem{ferlez2020shieldnn}
J.~Ferlez, M.~Elnaggar, Y.~Shoukry, and C.~Fleming, ``Shieldnn: A provably safe
  nn filter for unsafe nn controllers,'' \emph{arXiv preprint
  arXiv:2006.09564}, 2020.

\bibitem{robey2020learning}
A.~Robey, H.~Hu, L.~Lindemann, H.~Zhang, D.~V. Dimarogonas, S.~Tu, and
  N.~Matni, ``Learning control barrier functions from expert demonstrations,''
  \emph{arXiv preprint arXiv:2004.03315}, 2020.

\bibitem{Kha96}
H.~K. Khalil, \emph{Nonlinear Systems}, 2nd~ed.\hskip 1em plus 0.5em minus
  0.4em\relax Englewood Cliffs, NJ: Prentice-Hall, 1996.

\bibitem{goebel2009hybrid}
R.~Goebel, R.~G. Sanfelice, and A.~R. Teel, ``Hybrid dynamical systems,''
  \emph{IEEE Control Systems}, vol.~29, no.~2, pp. 28--93, 2009.

\bibitem{fazlyab2019efficient}
M.~Fazlyab, A.~Robey, H.~Hassani, M.~Morari, and G.~Pappas, ``Efficient and
  accurate estimation of lipschitz constants for deep neural networks,'' in
  \emph{Advances in Neural Information Processing Systems}, 2019, pp.
  11\,427--11\,438.

\bibitem{drake}
\BIBentryALTinterwordspacing
R.~Tedrake and the Drake Development~Team, ``Drake: Model-based design and
  verification for robotics,'' 2019. [Online]. Available:
  \url{https://drake.mit.edu}
\BIBentrySTDinterwordspacing

\bibitem{bradbury2020jax}
J.~Bradbury, R.~Frostig, P.~Hawkins, M.~J. Johnson, C.~Leary, D.~Maclaurin, and
  S.~Wanderman-Milne, ``Jax: composable transformations of python+ numpy
  programs, 2018,'' \emph{URL http://github. com/google/jax}, p.~18, 2020.

\bibitem{goswami1996limit}
A.~Goswami, B.~Espiau, and A.~Keramane, ``Limit cycles and their stability in a
  passive bipedal gait,'' in \emph{Proceedings of IEEE international conference
  on robotics and automation}, vol.~1.\hskip 1em plus 0.5em minus 0.4em\relax
  IEEE, 1996, pp. 246--251.

\bibitem{byl2008approximate}
K.~Byl and R.~Tedrake, ``Approximate optimal control of the compass gait on
  rough terrain,'' in \emph{2008 IEEE International Conference on Robotics and
  Automation}.\hskip 1em plus 0.5em minus 0.4em\relax IEEE, 2008, pp.
  1258--1263.

\bibitem{goswami1997limit}
A.~Goswami, B.~Espiau, and A.~Keramane, ``Limit cycles in a passive compass
  gait biped and passivity-mimicking control laws,'' \emph{Autonomous Robots},
  vol.~4, no.~3, pp. 273--286, 1997.

\end{thebibliography}

\newpage
\begin{appendix}

\section{Appendix A}
\subsection{Computing the Lipschitz constants $L_h$, $L_q^c$, and $L_q^c$}

 To verify \eqref{eq:lip1}, \eqref{eq:lip2_c}, and \eqref{eq:lip2_d}, the Lipschitz constant $L_h$ of the learned function $h(z)$ as well the Lipschitz constant $L_{\nabla h}$ of its gradient $\nabla h(z)$ need to be calculated first. The following discussion is mainly taken from \cite[Section 3.4.2]{robey2020learning}. When $\mathcal{H}$ consists of twice differentiable functions, as is the case when $h\in\mathcal{H}$ is parametrized by a Deep Neural Net with twice differentiable activation functions or a Reproducing  Kernel Hilbert Space, the following is shown to hold.

\emph{Reproducing  Kernel Hilbert Space: } In the case of random Fourier features with $l$ random features where $h(z) := \langle\phi(z),\theta\rangle$ with $\theta\in\Theta$ where $\Theta$ is a convex set and with $\phi:\mathbb{R}^{n_z} \to \mathbb{R}^l$ where
\begin{align*}
    \phi(z) = \sqrt{{2}/{l}} (\cos(\langle z,w_1\rangle + b_1), ..., \cos(\langle z,w_l \rangle + b_l)) \:,
\end{align*}
 an upper bound on the Lipschitz constant of $h(z)$ is provided as $\sqrt{2\sigma^2} (1 + \sqrt{n_z/l} + \sqrt{(2/l) \log(1/\delta)}) \|\theta\|_2$ with probability at least $1-\delta$ where $\sigma$ is as explained in \cite[Section 3.4.2]{robey2020learning}. An upper bound on the Lipschitz constant of $\nabla h(z)$ can be derived by the bound $\|\nabla^2 h(z)\| \leq 3\sqrt{2}\|\theta\|_\infty \sigma^2 (l+n_z+2\log(1/\delta))/\sqrt{l}$ that holds with probability at least $1-\delta$.

\emph{Deep Neural Net: } When $h(z)$ is a DNN, the problem of exactly computing the Lipschitz constant of $h(z)$ is known to be NP-hard.  Because most commonly-used activation functions $\phi$ are known to be 1-Lipschitz (e.g., ReLU, tanh, sigmoid), a naive upper bound on the Lipschitz constant of $h(z)$ is given by the product of the norms of the weight matrices; that is, $L_h \leq \prod_k\|W^k\|$.  However, this bound is known to be quite loose \cite{fazlyab2019efficient}.  Recently, the authors of \cite{fazlyab2019efficient} proposed a semidefinite-programming based approach to efficiently compute an accurate upper bound on $L_h$.  On the other hand, there are relatively few results that provide accurate upper bounds for the Lipschitz constant of the gradient of $h(z)$.  The only general method for computing an upper bound on $L_{\nabla h}$ is through post-hoc sampling.

Now, using these Lipschitz constants $L_h$ and $L_{\nabla h}$ of $h(z)$ and $\nabla h(z)$, respectively, it can be seen that the functions $q_c(z,u_c^i)$ and $q_d(z,u_d^i)$ in \eqref{eq:cons_3} and \eqref{eq:cons_4} are locally Lipschitz continuous in $z$ since $\nabla h(z)$, $f_c(z)$, $f_d(z)$, $g_c(z)$, $g_d(z)$, and $\alpha(h(z))$ are locally Lipschitz continuous and since function composition preserves Lipschitz continuity. Upper bounds of the Lipschitz constants $L_q^c$ and $L_q^d$ follow immediately.


\newpage

\section{Appendix B: Bouncing Ball}

The code for the bouncing ball case study can be found at \href{https://github.com/unstable-zeros/learning-hcbfs}{https://github.com/unstable-zeros/learning-hcbfs}.

\subsection{Generating and tracking the reference path}

The control goal of the bouncing ball case study is to track a desired reference path shown as dash-dotted blue lines in Fig.~\ref{fig:ball} (middle and right).
This reference path is obtained by dropping the ball from $(1,0)$ without applying any control input, i.e., $u_c:=0$,  until it hits the ground.
The discrete control input $u_d$ is set to $1.0$ when this happens.
Reversing the velocity in sign while keeping the position unchanged gives the reference path in the right half plane.

\noindent In order to track the reference path, we design a linear error-feedback control law $u_{c,\text{nom}} = K (z(t,j)-z_{\text{ref}}(t))$ with $K = \begin{bmatrix} 10 & 5.48 \end{bmatrix}^T$ via solving the Riccati equation.
The reference state $z_{\text{ref}}(t)$ is selected as the state on the reference path that is the closet to $z(t,j)$ in Euclidean distance.
We set the nominal discrete input to be $u_{d,\text{nom}} = 1.2$ such that it amplifies the velocities after collisions and thus is more likely to cause constraint violation.
Note that neither the reference path nor the closed-loop trajectory obtained by applying the nominal controllers $u_{c,\text{nom}}$ and $u_{d,\text{nom}}$ complies with the geometric safe set $\mathcal{S}$.

\subsection{Training data}
To obtain training data during flows, we use the above reference controller $u_{c,\text{nom}}$ together with the analytical HCBF in Example \ref{ex:2}, which, per solving the CBF-QP problem \cite{ames2014control}, gives a safe controller $u_{c}$ during flow. The CBF-QP basically consists of solving a convex quadratic program with decision variable $u_{c}$, cost function $\|u_{c}-u_{c,\text{nom}}\|$, and the constraint $\langle \nabla h(z), f_c(z)+g_c(z)u_{c}\rangle \ge -\alpha(h(z))$. The safe control input $u_{d}$ used to obtain training data  during jumps is essentially a thresholding function and can be found analytically as mentioned in Example \ref{ex:2}. Based on these safe controllers $u_{c}$ and $u_{d}$, we obtain data-sets $Z^c_{\text{safe}}$ and $Z^d_{\text{safe}}$ containing 5000 safe states $z^i$ and associated expert demonstrations $u^i_c$ and $u^i_d$, which are shown as the green and magenta dots in the left subplot of Figure \ref{fig:ball}.

\subsection{Hyperparameters for training the neural network}
We train the neural network for $1.5 \times 10^3$ epochs using the loss given by unconstrained relaxation of Problem \eqref{eq:opt} in Section 3.
The hyperparameters, shown in Table \ref{tab:nn-params-bb}, are found by grid search.

\begin{table}[H]
    \centering
    \begin{tabular}{|c|c|} \hline
         \thead{Parameter name} & \thead{Value}  \\ \hline
         $\lambda_{\mathrm{s}}$ & 4.0 \\ \hline
         $\lambda_{\mathrm{u}}$ & 5.0 \\ \hline
         $\lambda_{\mathrm{d}}$ & 1.0 \\ \hline
         $\lambda_{\mathrm{c}}$ & 1.0 \\ \hline
         $\gamma_{\text{safe}}$ & 0.0025 \\ \hline
         $\gamma_{\text{unsafe}} $ & 0.075 \\ \hline
         $\gamma_{\text{dyn}}^c$ & 0.055 \\[0.05cm] \hline
         $\gamma_{\text{dyn}}^d$ & 0.055 \\ \hline
    \end{tabular}
    \vspace{3pt}
    \caption{Hyperparameters used for training the neural network based HCBF for the bouncing ball.}
    \label{tab:nn-params-bb}
\end{table}
\vspace{-15pt}

\subsection{Visualization of the learned HCBF}
\begin{figure}[H]
\centering
\includegraphics[scale=0.45]{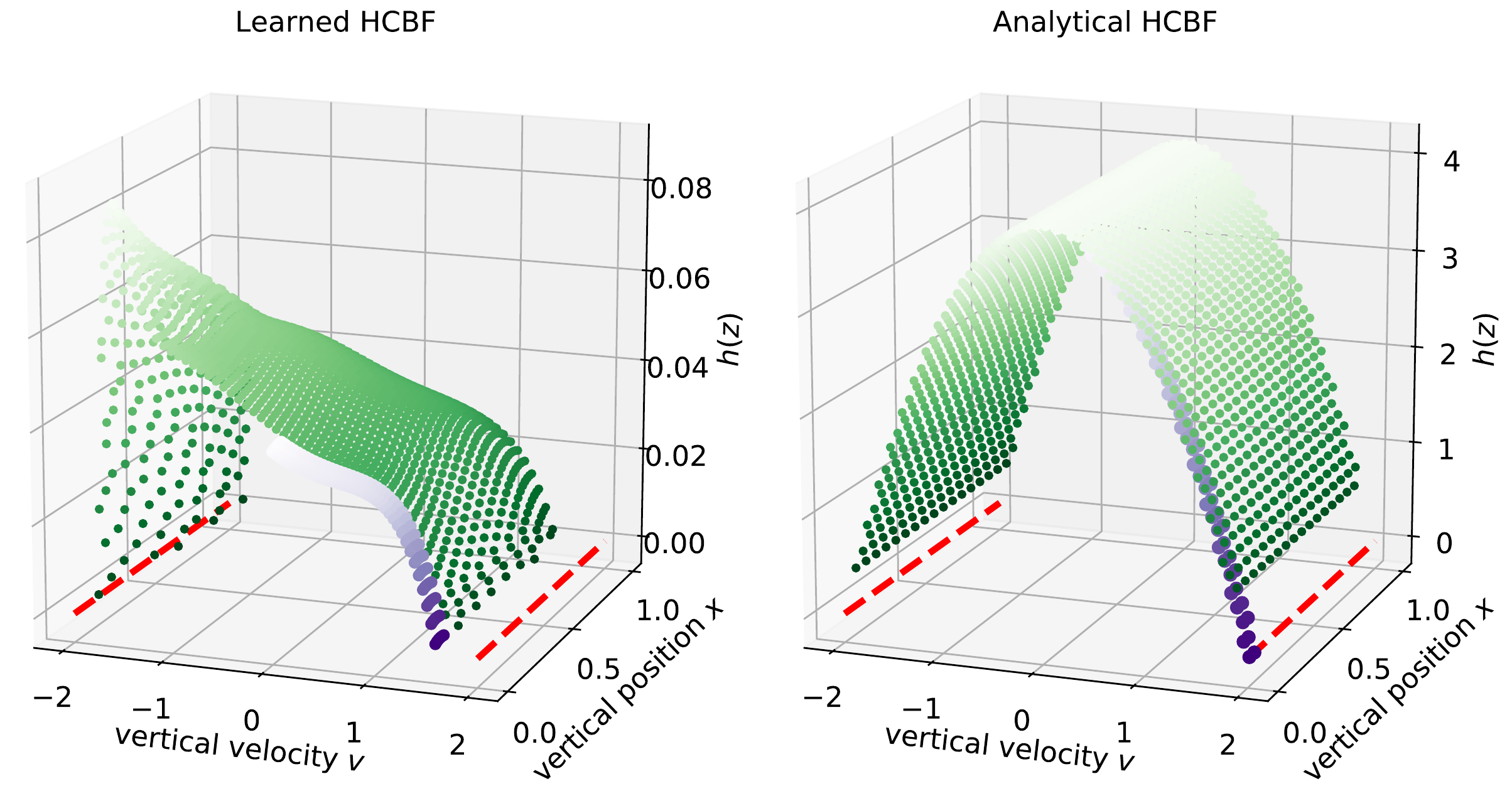}
\caption{Left: Level sets of the learned HCBF. Right: Level sets of the analytical HCBF. For both plots, green dots indicate where $h(z) \geq 0$ on $C\cup D$, while the purple dots additionally indicate where $h(z) \geq 0$ on  $\{ z\in C \mid \ x=0 \ , \ 0 \leq v \leq 2 \}$. Deeper color indicates a lower value of $h(z)$.}
\label{fig:HCBF_3D}
\end{figure}

We visualize the learned and analytical HCBFs in $(v,x,h(z))$-space in Figure \ref{fig:HCBF_3D} by plotting their level sets. The green dots indicate where $h(z) \geq 0$ on $C\cup D$, while the purple dots additionally indicate where $h(z) \geq 0$ on  $\{ z\in C \mid \ x=0 \ , \ 0 \leq v \leq 2 \}$, i.e. the possible set of states after an impact with the ground.

\subsection{Validness of the learned HCBF}
We performed post-verification of the learned HCBF satisfying Propositions \ref{prop:1}, \ref{cor:1} and \ref{prop:2}, which serve as sufficient conditions for learning a valid HCBF as stated in Theorem \ref{thm:2}.
Since all of our data sets were gridded evenly on the state space, densities of the $\epsilon$-nets are equal to the gridding resolutions, which are $\bar{\epsilon} = 0.01$ and $\epsilon_c = \epsilon_d = 0.02$.
The local Lipschitz constants $L_h(\cdot), L_q^c(\cdot), L_q^d(\cdot)$ were approximated using the norm of their gradients, e.g. $L_h(z) \approx \| \nabla h(z) \|_2$, which gives a tight estimation of the Lipschitz constants due to our dense gridding scheme.
The percentages of data points that satisfy those conditions are summarized in Table \ref{tab:sat_rate_bb}.
\begin{table}[H]
    \centering
    \begin{tabular}{|c|c|c|} \hline
         \thead{Condition} & \thead{Involved datasets} & \thead{Satisfaction rate}  \\ \hline
                  \eqref{eq:cons_1} & $Z_\text{safe}^c$, $Z_\text{safe}^d$ & $95.38\%$   \\ \hline
         \eqref{eq:cons_2} & $Z_N$ & $96.6\%$   \\ \hline
         \eqref{eq:cons_3} & $Z_\text{safe}^c$, $Z_\text{dyn}^c$ & $100\%$  \\ \hline
         \eqref{eq:cons_4} & $Z_\text{safe}^d$, $Z_\text{dyn}^d$ & $100\%$  \\ \hline
         Proposition \ref{prop:1} & $Z_N$ & $100\%$  \\ \hline
         Proposition \ref{cor:1} & $Z_\text{safe}^c$, $Z_\text{safe}^d$ & $98.81\%$   \\ \hline
         Proposition \ref{prop:2} & $Z_\text{safe}^c$, $Z_\text{safe}^d$, $Z_\text{dyn}^c$, $Z_\text{dyn}^d$ & $98.74\%$  \\ \hline
    \end{tabular}
    \vspace{3pt}
    \caption{Satisfaction rates of sufficient conditions for learning a valid HCBF for the bouncing ball.}
    \label{tab:sat_rate_bb}
\end{table}

\newpage

\section{Appendix C: Compass Gait Walker}
The code for the compass gait walker case study can be found at \href{https://github.com/unstable-zeros/learning-hcbfs}{https://github.com/unstable-zeros/learning-hcbfs}. 

\subsection{Sampling boundary points}

\begin{algorithm}
    \centering
    \begin{algorithmic}[1]
        \Statex \textbf{Input: } Tuple of $n_z$-dimensional vectors $Z := Z_\text{safe}^c\cup Z_\text{safe}^d$, minimum number of neighbors $N\in\mathbb{Z}_+$, neighbor threshold $\eta > 0$
        \Statex \textbf{Output: } Binary vector $\{y_1, \dots, y_n\} \subset \{0, 1\}^n$
        \Statex
        \State $M \gets compute\_pairwise\_dists(Z)$ \quad\# compute pairwise distances between vectors in $Z$
        \State $\tilde{M}_{i,j} \gets \mathbbm{1}(M_{i,j} \leq \eta)$ for $(i,j)\in \{1,\hdots,n\} \times \{1,\hdots,n\}$ \quad\# threshold $M$ by $\eta$
        \State $\tilde{y} \gets sum(\tilde{M}, \text{axis}=1)$ \quad\# sum across the rows of $\tilde{M}$
        \State $y \gets \mathbbm{1}(\tilde{y} \leq N)$ \quad\# threshold $\tilde{y}$ by $N$
    \end{algorithmic}
    \caption{Neighbor-based Unsafe Trajectory Sampling (NUTS)}
    \label{alg:mr-peanut}
\end{algorithm}

The approach we have described crucially relies on being able to sample from ``unsafe'' regions of state space, i.e. to sample states $z^i \in Z_N$.  For low-dimensional state spaces, it is often possible to take advantage of the underlying geometry of the hybrid system to sample from this region as for instance in the bouncing ball case study.  However, for higher-dimensional state spaces such as the compass gait walker, it may not be possible to easily leverage the underlying geometry to obtain these states.  

To collect unsafe states $z^i \in Z_N$ for dynamical or hybrid systems with high dimensional state spaces, we propose an algorithm for obtaining these unsafe states by sampling the expert trajectories.  The pseudocode for this algorithm is given in Algorithm \ref{alg:mr-peanut}.  This algorithm takes two input parameters: a positive integer $N\in\mathbb{Z}_+$ and a small nonnegative constant $\eta>0$.  In line 1, we first compute the pairwise distances between each of the $n:=N_c+N_d$ states in our expert trajectories $Z_\text{safe}^c\cup Z_\text{safe}^d$; the result is a symmetric $n\times n$ matrix $M$, where the element at position $(i,j)$ in $M$ represents the pairwise distance between states $z^i$ and $z^j$.  Next, we threshold $M$ so that all values in $M$ that are greater than $\eta$ are set to zero in $\tilde M$ and all values in $M$ that are less than $\eta$ are set to one. In line 2, we use the specific notation $\mathbbm{1}(M_{i,j} \leq \eta) := 1$ if $M_{i,j} \leq \eta$  and $\mathbbm{1}(M_{i,j} \leq \eta) :=0$ otherwise. Intuitively, the idea in this step is to identify the neighbors of each data point.  More specifically, the $i$th row of $\tilde{M}$ corresponds to a state $z^i$; then the indices $j$ in this row with $M_{ij} < \eta$ correspond to states $z^j$ which we deem neighbors of $z^i$.  In line 3, we sum along the rows of $\tilde{M}$ to create the vector $\tilde y \in\mathbb{R}^n$, which counts the number of neighbors for each state $z^i$ for $i\in \{1, 2, \dots, n\}$.  Finally in line 4, we threshold this vector $\tilde{y}$ based on the minimum number of neighbors $N$. Note here that $\mathbbm{1}(\tilde{y} \leq N)$ operates element-wise.  The result is a binary vector $y\in\mathbb{R}^n$, where the indices $i$ that are set to one correspond to states $z^i$ that the algorithm identifies as boundary points.  In this way, the total number of boundary points identified by the algorithm is $\sum_{i=1}^n y_i$.

\begin{figure}
    \centering
    \begin{subfigure}{0.31\textwidth}
        \includegraphics[width=\textwidth]{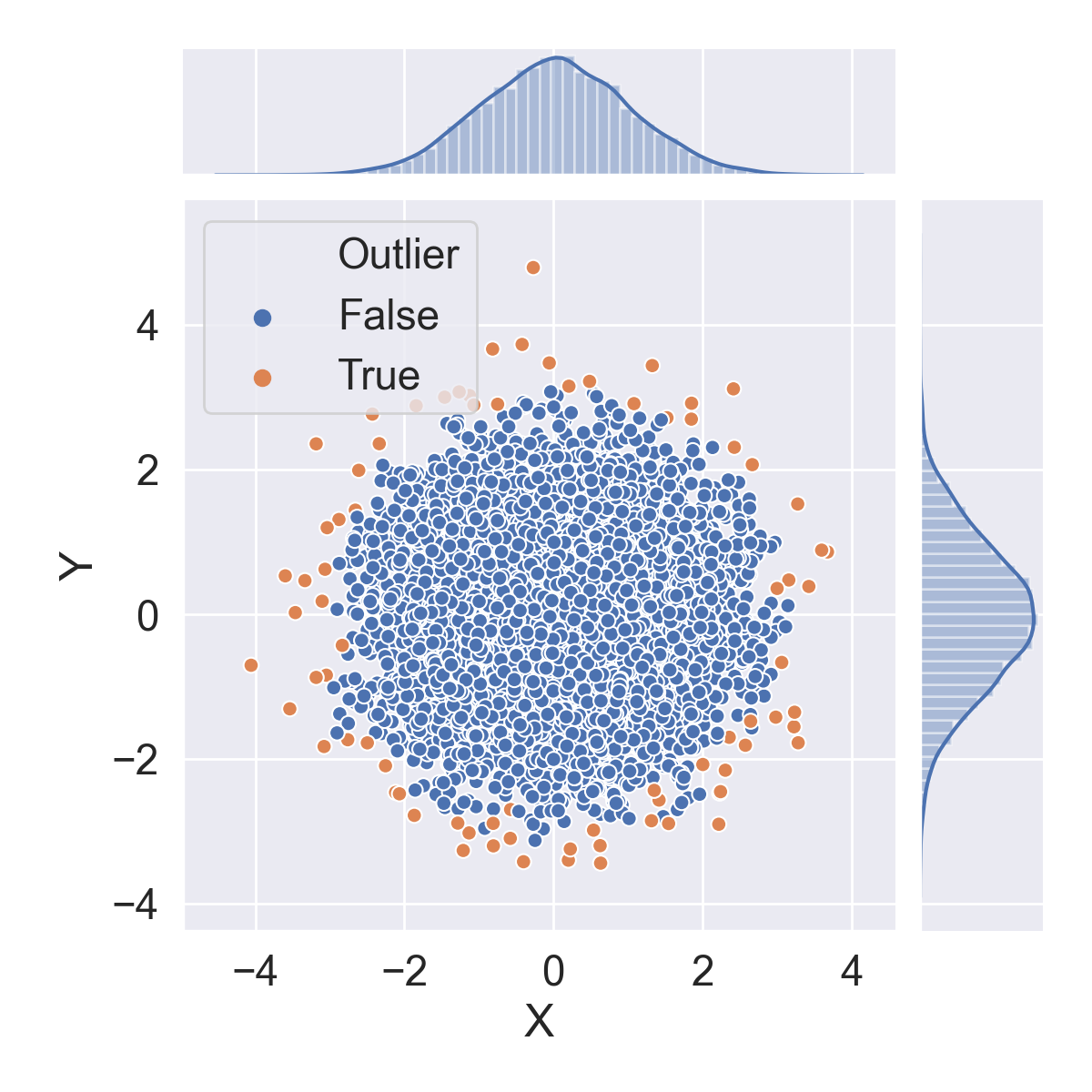}
        \caption{We generate 20000 data points according to $z\sim \mathcal{N}(0, I)$.  The marginals of the distribution are shown above and to the right of the plot.}
        \label{fig:one-gaussian}
    \end{subfigure} \quad
    \begin{subfigure}{0.31\textwidth}
        \includegraphics[width=\textwidth]{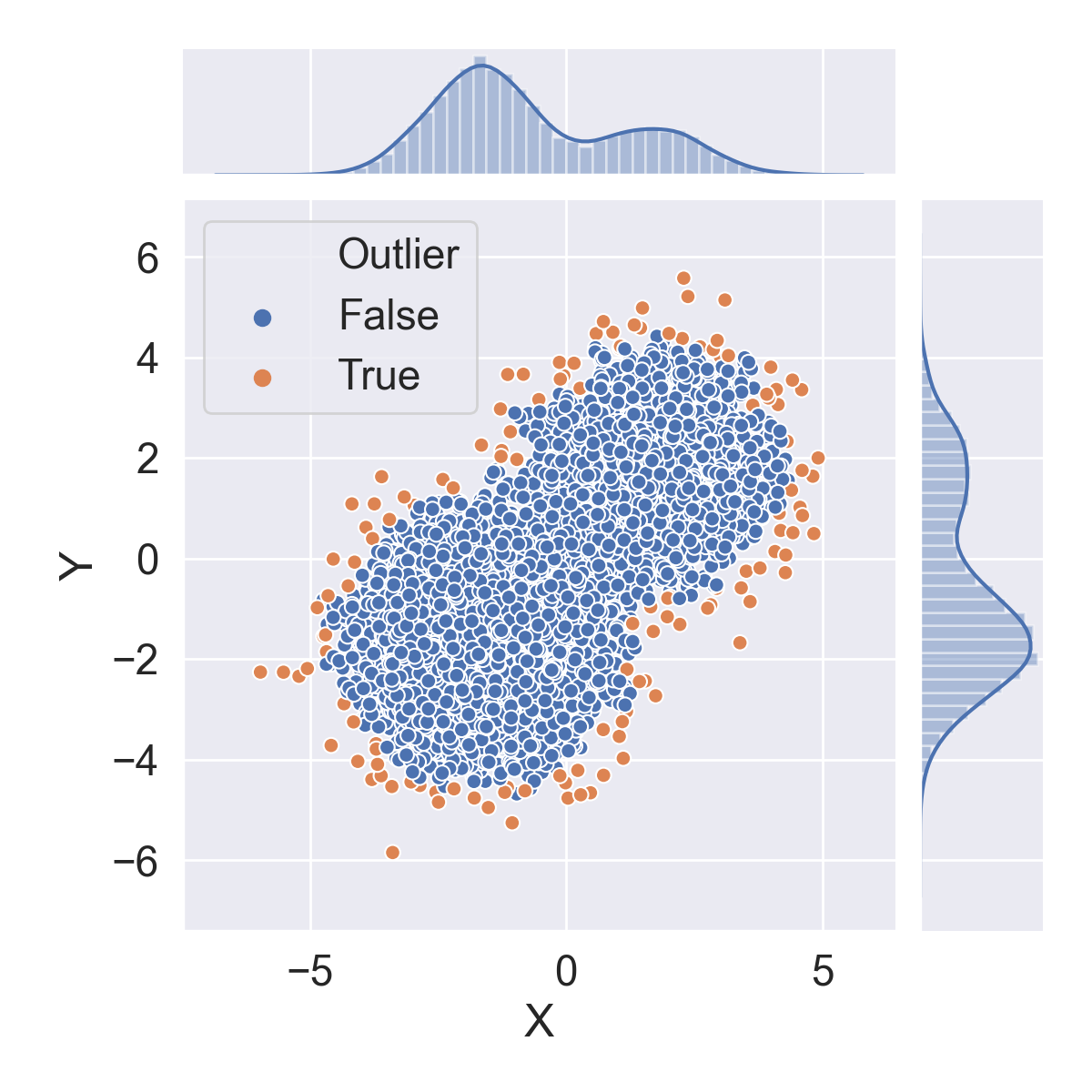}
        \caption{We mix two Gaussians.  In this case, we generate 20000 data points $z$ according to $z|y \sim\mathcal{N}(1.7y\mathbbm{1}, I)$ with $x\sim \text{Bernoulli}(1/2)$ and $y = 2x-1$}
        \label{fig:mixture-gaussians}
    \end{subfigure} \quad
    \begin{subfigure}{0.31\textwidth}
        \includegraphics[width=\textwidth]{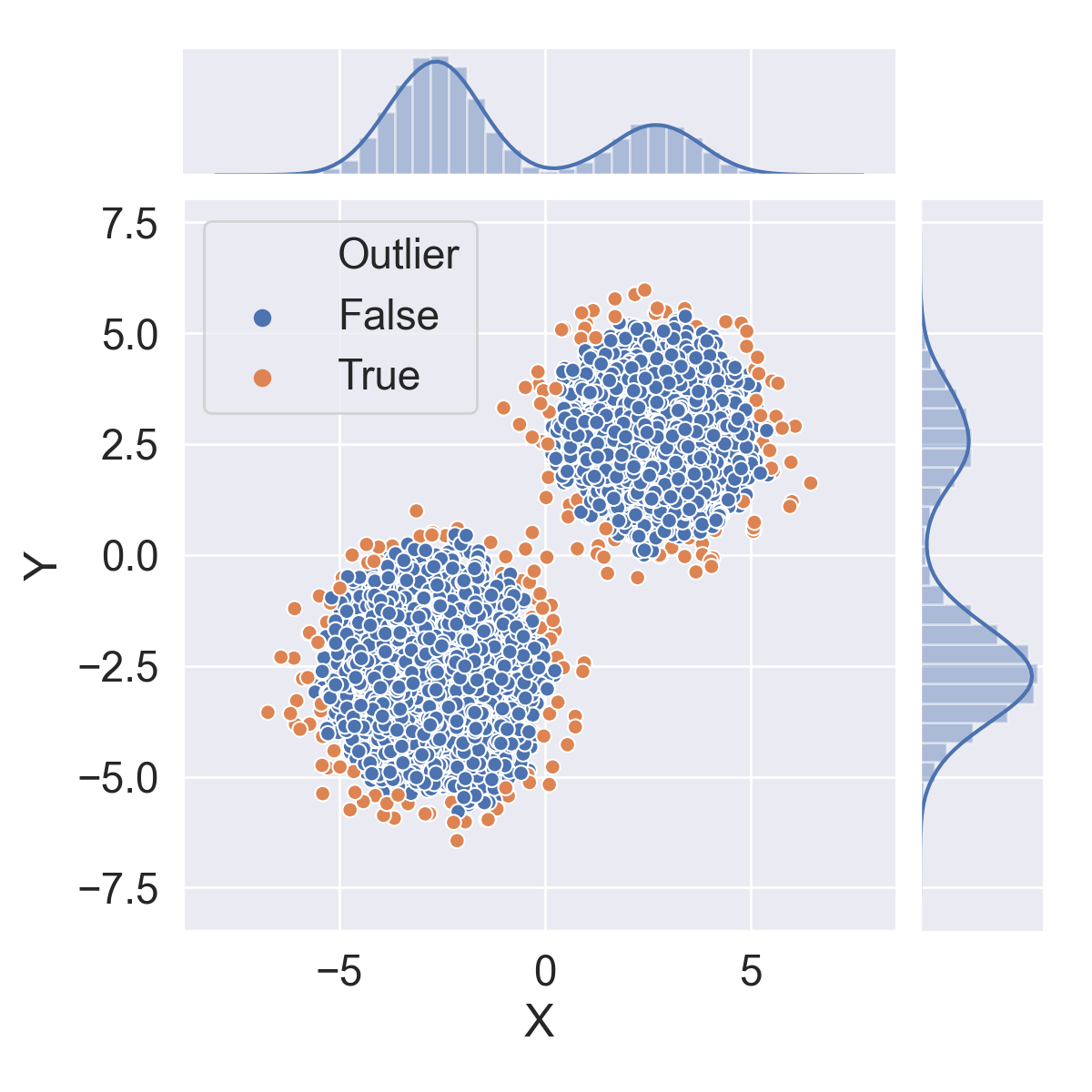}
        \caption{Again, we mix two Gaussians.  In this case, we generate 20000 data points $z$ according to $z|y \sim\mathcal{N}(2.7y\mathbbm{1}, I)$ with $x\sim \text{Bernoulli}(1/2)$ and $y=2x-1$}
        \label{fig:separate-gaussians}
    \end{subfigure}
    \caption{We run Algorithm 1 on data generated from two-dimensional Gaussian distributions with $\eta=0.3$ and $N=5$.  All boundary points are marked in orange, and the non-boundary points are marked in blue.}
    \label{fig:separate-gaussians}
\end{figure}

\begin{figure}
    \centering
    \includegraphics[width=0.9\textwidth]{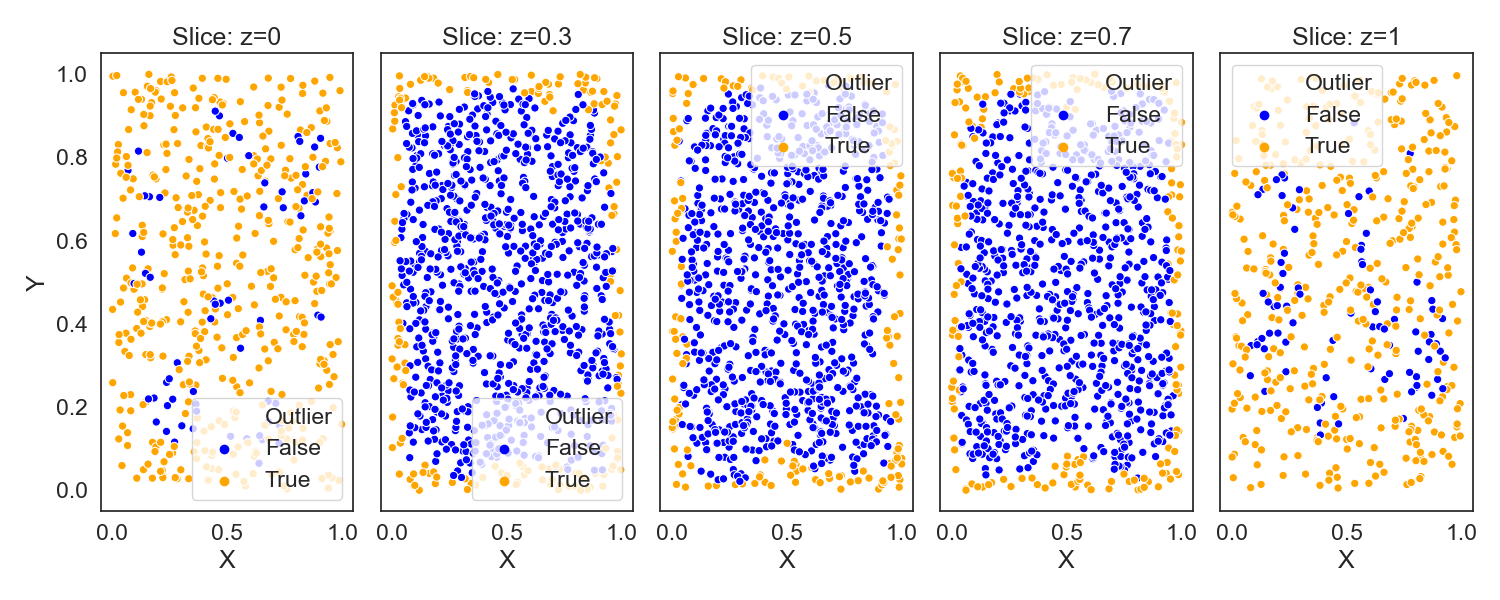}
    \includegraphics[width=0.9\textwidth]{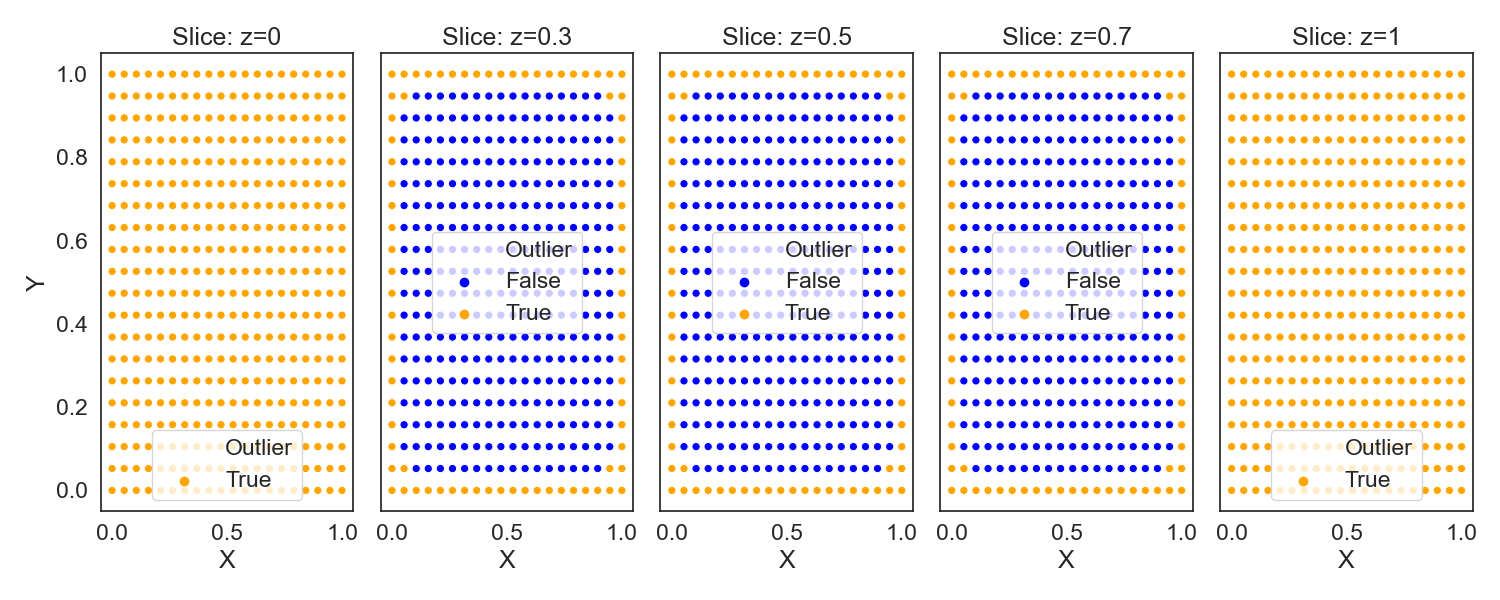}
    \caption{(Top) We uniformly sample 1000 points $(x,y,z)\in\mathbb{R}^3$ in the unit cube.  We then run Algorithm \ref{alg:mr-peanut} with $\eta = 0.13$ and $N=70$ and show slices of this state space $\{(x,y,z) | z_0 - 0.05 \leq z \leq z_0 + 0.05\}$ for $z_0\in\{0, 0.3, 0.5, 0.7, 1\}$.  Notice that on the boundary slices $z_0 = 0$ and $z_0 = 1$, the algorithm identifies the majority of the slice as a boundary points.  On the other hand, for the slices in the middle of the unit cube, generally only the outline of the slice contains boundary points.  In this way, the algorithm successfully identifies the boundary of the cube.  (Bottom) We repeat the same experiment as the top panel, except in this case we grid the unit cube with 8000 points and rerun the algorithm with the same parameters.  Notice that the boundary of the cube is identified exactly in this case.}
    \label{fig:outlier-grids}
\end{figure}

\begin{figure}
    \centering
    \includegraphics[width=\textwidth]{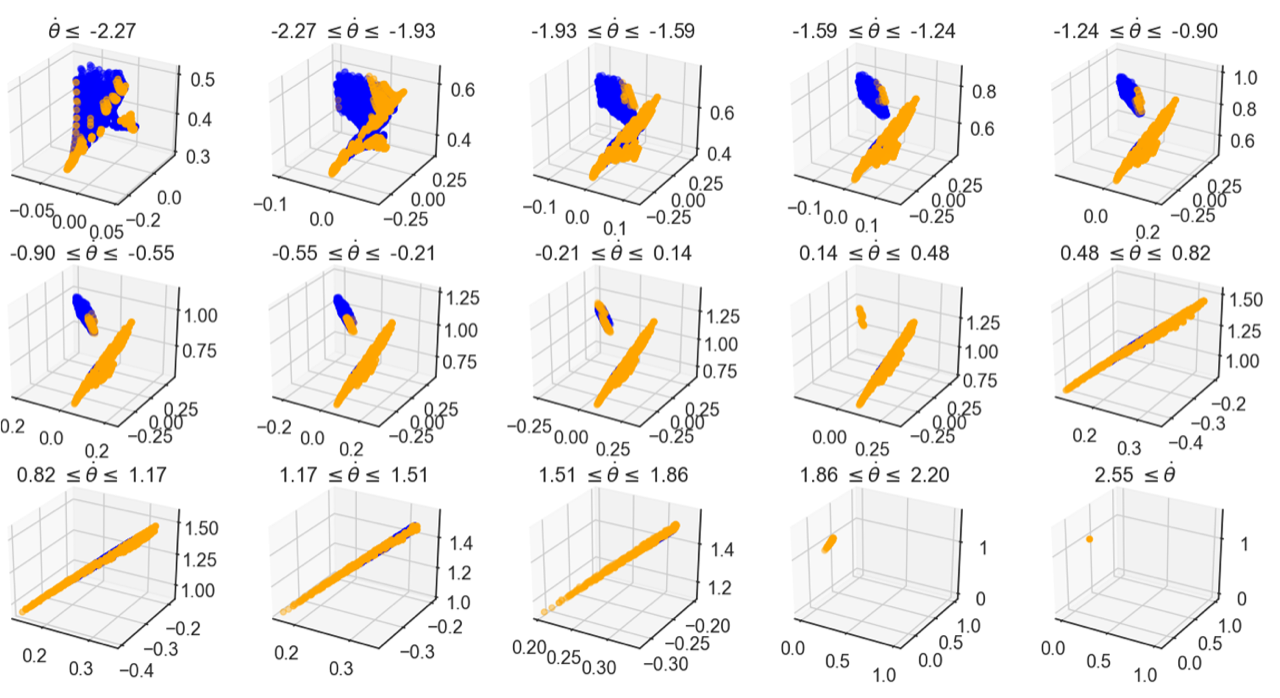}
    \caption{We show the result of running Algorithm \ref{alg:mr-peanut} with $\eta=0.035$ and $N=50$ on 10 rollouts for the compass gait walker from different initial conditions sampled in the neighborhood of a point $[0.0, 0.0, 0.4, -2.0]$ on the passive limit cycle.  As the state space is four-dimensional, we bin the states in these rollouts according to the angular velocity of the swing leg and plot the other components of the state in each subplot.  In particular, the $x, y,$ and $z$ axes of each plot are angle of the stance leg, the angle of the swing leg, and the angular velocity of the stance leg respectively.  Boundary points are shown in orange, whereas the interior points are shown in blue.}
    \label{fig:cg-bdy}
\end{figure}

Before demonstrating how we used this algorithm in the compass gait case study, we provide some simple examples to illustrate the efficacy of this algorithm toward identifying the boundaries of different sets of points.  In Figure \ref{fig:separate-gaussians}, we show the result of running Algorithm \ref{alg:mr-peanut} on mixtures of two Gaussians.  In Figure \ref{fig:outlier-grids}, we show a three-dimensional example in the unit cube.  Finally, we show the result of running Algorithm \ref{alg:mr-peanut} on 100,000 states taken from the expert compass gait walker in Figure \ref{fig:cg-bdy}. 

\subsection{Implementation details}

\begin{table}[]
    \centering
    \begin{tabular}{|c|c|} \hline
         \thead{Parameter name} & \thead{Value} \\ \hline
         Hip mass $m_H$ & 10.0 kg \\ \hline
         Leg mass $m_L$ & 5.0 kg \\ \hline
         Length leg $\ell$ & 1.0 m \\ \hline
         Leg center of mass $a$ & 0.5 m \\ \hline
         Gravity $g$ & 9.81 m/$\text{s}^2$ \\ \hline
         Slope angle $\gamma$ & 0.0525 radians \\ \hline
    \end{tabular}
    \vspace{3pt}
    \caption{Parameters used for the compass gait walker.}
    \label{tab:compass-gait params}
\end{table}

Our implementation of the compass gait walker relies heavily on the C++ implementation in the Drake package \cite{drake}.  In particular, we re-implement the compass gait walker in Python with Jax \cite{bradbury2020jax}: our implementation is publicly available on the project \href{https://github.com/unstable-zeros/learning-hcbfs}{Github repository}.  We use the same default settings for the compass gait walker as are used in the Drake implementation; these values are detailed in Table \ref{tab:compass-gait params}. 

The hyperparameters used for training a HCBF for the compass gait walker are given in Table \ref{tab:nn-params-cg}.  These hyperparameters were chosen via grid search.

\begin{table}[]
    \centering
    \begin{tabular}{|c|c|} \hline
         \thead{Parameter name} & \thead{Value}  \\ \hline
         $\lambda_{\mathrm{s}}$ & 5.0 \\ \hline
         $\lambda_{\mathrm{u}}$ & 5.0 \\ \hline
         $\lambda_{\mathrm{d}}$ & 0.5 \\ \hline
         $\lambda_{\mathrm{c}}$ & 0.5 \\ \hline
         $\gamma_{\text{safe}}$ & 0.3 \\ \hline
         $\gamma_{\text{unsafe}} $ & 0.3 \\ \hline
         $\gamma_{\text{dyn}}^c$ & 0.05 \\ \hline
         $\gamma_{\text{dyn}}^d$ & 0.05 \\ \hline
    \end{tabular}
    \vspace{3pt}
    \caption{Hyperparameters used for training the neural network based HCBF for the compass gait.}
    \label{tab:nn-params-cg}
\end{table}

\subsection{Collecting expert trajectories}

To collect expert trajectories, we use the energy-based controller described in eq.\ (24) in \cite{goswami1997limit}.  This controller applies actuation to the hip joint, leaving the ankle joints unactuated.  In particular, this controller takes the following form
\begin{align*}
    u_H = - \frac{\lambda(E - E^*)}{\dot{\theta}_{\text{stance}} - \dot{\theta}_{\text{swing}}}
\end{align*}
where we set $\lambda:=0.3$ and where $E^*$ is the reference energy from the passive limit cycle, and $E$ is the current total mechanical energy of the compass gait walker.  Throughout, we use a reference energy of $E^* = 153.244 J$ as suggested by \cite{goswami1997limit}.

As described in the main text, we collect expert trajectories by fixing the initial condition of the left leg to $[\theta_{\text{stance}}, \dot{\theta}_{\text{stance}}] = [0, 0.4]$ and varying the initial condition of the right foot $[\theta_{\text{swing}}, \dot{\theta}_{\text{swing}}] = [0 + u_1, 2.0 + u_2]$ where $u_1\sim\text{Uniform}([-0.2, 0.2])$ and $u_2 \sim \text{Uniform}([-0.5, 0.5])$. In Figure \ref{fig:init-conds-compass-gait} (right), we show the successful right foot initial conditions for the energy-based controller corresponding to this initial condition. In particular, to train the HCBF, we collected 500 rollouts using this scheme.  For each rollout, we used a horizon of $T = 750$ steps and a time interval of $\Delta t = 0.01$.

\end{appendix}

\end{document}